\documentclass[useAMS,usenatbib]{mn2e}
\usepackage{graphicx}
\usepackage{amssymb}
\usepackage{lscape}
\usepackage{rotating}
\usepackage{subfigure}

\def\deg{\ifmmode^\circ\else$^\circ$\fi}

\def\Q{\ifmmode\mathcal{Q}\else$\mathcal{Q}$\fi}
\def\Mach{\ifmmode\mathcal{M}\else$\mathcal{M}$\fi}

\title[{\it Spitzer} IRAC study of M8]
{ A study of the massive star forming region M8 using {\it Spitzer} IRAC images}

\author[L.K. Dewangan \& B.G. Anandarao]
{Lokesh Kumar Dewangan$^{1}$\thanks{lokeshd@prl.res.in}, \& B.G. Anandarao$^{1}$\thanks{anand@prl.res.in}\\
$^1$Astronomy $\&$ Astrophysics Division, Physical Research Laboratory, Navrangpura, Ahmedabad 380 009, India.\\}
\begin{document}

\date{ }

\pagerange{\pageref{firstpage}--\pageref{lastpage}} \pubyear{2009}

\maketitle

\label{firstpage}

\begin{abstract}{\it Spitzer} IRAC images (3.6, 4.5, 5.8 and 8.0$\mu$m) and photometry 
of the star forming region M8 are presented. IRAC photometry reveals ongoing star 
formation in the M8 complex, with 64 Class 0/I and 168
Class II sources identified in several locations in the vicinity of sub-mm gas 
cores/clumps. Nearly 60\% of these YSOs occur in about 7 small clusters. 
The spatial surface density of the clustered YSOs is determined to be about 10-20 YSOs/pc$^{2}$. 
Fresh star formation by the process of ``collect and collapse'' 
might have been triggered by the expanding HII regions and winds from massive stars.  
IRAC ratio images are generated and studied in order to identify possible diagnostic emission regions in M8. 
The image of 4.5$\mu$m/8.0$\mu$m reveals Br$\alpha$ counterpart of the optical Hourglass HII region, 
while the ratio 8.0$\mu$m/4.5$\mu$m indicates PAH emission in a cavity-like structure to the east of the Hourglass. 
The ratio maps of  3.6$\mu$m/4.5$\mu$m, 5.8$\mu$m/4.5$\mu$m and 8.0$\mu$m/4.5$\mu$m 
seem to identify PAH emission regions in the sharp ridges and filamentary structures 
seen East to West and NE to SW in M8 complex. 

\end{abstract}

\begin{keywords}
stars: formation -- stars: pre-main-sequence --  stars: winds and outflows 
-- infrared: ISM -- ISM: HII Regions -- ISM: Individual: M8
\end{keywords}

\section{Introduction}
\label{sec:intro}
Messier 8 (M8), the Lagoon Nebula (or NGC 6523) is a well known galactic HII 
region (comprehensively reviewed in \citet{Tothill08}, and the references therein) situated at a 
distance of 1.25 kpc \citep{Arias06} in Sagittarius-Carina 
spiral arm of the Galaxy. The core of M8 contains a spectacular 
blister-type HII region, called the Hourglass nebula, ionized by the O7.5 V 
star Herschel 36 (Her 36) \citep{Woolf61}. The HII region is embedded within a 
giant molecular cloud that extends eastwards to the young star cluster NGC 6530 
of age 2 $\times$ 10${^6}$ yrs \citep{Lada76}. 
While Her 36 is responsible for the Hourglass and the ionised bubble surrounding it, 
the other early type stars in M8 complex, 9 Sgr (O4 V(f)) and HD165052(O6.5V+O7.5V) 
are believed to account for the ionised regions east of the 
central core/bubble \citep{Goudis76,Lada76,Lynds82,Woodward86}. 
\citet{Allen86} discovered a few near-IR sources in the vicinity of Her 36,
designated by \citet{Woodward90} as KS 1 to KS 5. 
From a high resolution near-IR study, \citet{Arias06, Arias07} found the existence of 
a very young star cluster around Her 36, having an age of $\sim$ 10${^6}$ yrs. 
\citet{Barba07} discovered a number of HH objects in M8, which confirms by implication, the 
existence of very young stars undergoing the accretion phase of formation.
Narrow-band imaging by the Hubble Space Telescope (HST) 
revealed the presence of proplyds in the neighbourhood of Her 36 
\citep{Stecklum98}. The core of M8 is detected by the 
Mid-course Space Experiment (MSX) in mid-infrared as a luminous 
extended source \citep{Crowther03}. Color composite map of M8 in mid-infrared 
bands of {\it Spitzer} Infrared Array Camera (IRAC) 
shows a ridge extending in east-west direction to the south-east of the Hourglass \citep{Tothill08}.
There appear also a number of filamentary structures extending in the NE-SW direction to 
the east of Her 36 or the Hourglass. 
Among the new compact star forming regions in the M8 complex, M8E stands out
with its compact HII region powered by an early B type 
star (M8E-Radio) \citep{Lada76,Wright77,Brand78,Mitchell91,Linz09}.
\citet{White97} discovered very intense CO line emission 
in mm and sub-mm wavelength regions from the central core of M8. 
Later, from a larger survey in mm and sub-mm continuum and CO lines, 
\citet{Tothill02} found bright rims and dark lanes stretching in the east-west direction. 
\citet{White97} found from CO (J=3--2) 
line observations a loose bipolar structure extending NW-SE from Her 36; while \citet{Stecklum95} 
found the presence of a jet-like object very close to Her 36. These observations provide 
evidence for outflow activity around Her 36 region. 
As for the spectral diagnostics in the infrared region, \citet{Woodward86} detected Br$\gamma$ 
(2.17 $\mu$m), Br$\alpha$ (4.05 $\mu$m) as well as Pf$\epsilon$ (3.03 $\mu$m) 
and the 3.28 $\mu$m polycyclic aromatic hydrocarbons (PAH) 
emission to the west of Her 36. \citet{Burton02} observed 
H$_{2}$ 1--0 S(1) line at 2.12 $\mu$m near the Hourglass/Her 36 region attributed primarily to 
shock-excited molecular gas; but UV excitation can not be ruled out. 

The M8 region seems to be quite complex and very interesting, 
owing to the presence of stellar winds and expanding HII region 
bubbles from massive stars, which can trigger fresh star formation 
by sweeping up and compressing the local dense interstellar matter. 
The afore-mentioned near-IR surveys studied the regions around Her 36 and NGC 6530, 
leaving the ridge regions far-east and south-east of Her 36 relatively under-explored
in near-IR (except 2MASS survey) and mid-IR regions (except MSX survey). 
In view of this, we wish 
to examine in detail the extended region of M8 in the near-/mid-infrared images provided
by {\it Spitzer}, which are so far not looked into. 

{\it Spitzer} IRAC provides an opportunity with an unprecedented 
high spatial resolution in thermal infrared wavelength regime that is very useful in 
identifying embedded sources in massive star forming regions. IRAC has four wavelength 
bands (with $\lambda_{eff}/\Delta\lambda$, 3.55/0.75, 4.49/1.0, 5.73/1.43 and 7.87/2.91 $\mu$m) 
which include molecular emissions such as those from H$_{2}$ and PAH molecules, 
as well as recombination lines from hydrogen.
The aims of the present study are to identify the embedded sources using the IRAC bands
in order to classify the different stages of their evolution; to use ratio maps of the 
four IRAC bands in order to identify possible H$_{2}$, PAH or H emission regions,  
following the suggestions of \citet{Smith05} and \citet{Povich07}.

In Section 2, we describe the data used for the present study and the analysis tasks utilised. 
Section 3 presents the results and discussion on {\it Spitzer} IRAC photometry of embedded 
sources associated with M8 complex. In this section, we also present the results 
and discussion on the ratio maps. In Section 4, we give the conclusions.

\section{{\it Spitzer} IRAC Data on Messier 8 and Data Reduction}
\label{sec:data}
The {\it Spitzer} Space Telescope IRAC observation of M8 were 
obtained by Spitzer Science Center (SSC) on 16 September 2005 and Basic Calibrated Data (BCDs) 
images were processed by SSC using software version S14.0.0 for all four 
bands (see \citet{Fazio04}, for details on the IRAC instrument).
The observations relevant for M8 region were taken in High Dynamic Range (HDR) 
mode with 12s integration time in all filters. These observations were a part of the project entitled, 
``Spitzer Follow-up of HST Observations of Star Formation in H II Regions" 
(Program id 20726; PI: Jeff Hester).
The IRAC archival images of M8 were obtained by us on 8 April 2009 from the Spitzer public 
archive, using `leopard' software. The BCD images were processed for `jailbar' removal, 
saturation and `muxbleed' correction before making the final mosaic using Mopex
and IDL softwares \citep{Makovoz05}.
A pixel ratio (defined as the ratio of the area formed by the 
original pixel scale, 1.22 arcsec/px, to that 
of the mosaiced pixel scale) of 2 was adopted for making the mosaic 
(which gives a mosaic pixel scale of 0.86 arcsec/pixel) \footnote[1]{see 
http://ssc.spitzer.caltech.edu/postbcd/doc/mosaiker.pdf}. Using these procedures, a total number of 
320 BCD images of 5.2 $\times$ 5.2 arcmin$^{2}$ were mosaiced to 
make a final image of 42.5 $\times$ 30.0 arcmin$^{2}$ in each of the four bands. 
Aperture photometry was performed on the mosaic with 2.8 pixel aperture and sky 
annuli of 2.8 and 8.5 pixels using APPHOT task in IRAF package. The zero points for these apertures (including
aperture corrections) are, 17.08, 17.30, 16.70 and 15.88 mag  for the 3.6, 4.5, 5.8, 8.0 $\mu$m bands, 
here onwards called as Ch1, Ch2, Ch3 and Ch4 respectively (see \citet{Reach05}). 
The photometric uncertainties 
vary between 0.01 to 0.25 for the four channels, with Ch3 and Ch4 on the higher side.

We have looked into the 2MASS archives \citep{Skrutskie06} as well as published literature 
for JHK photometric data on sources 
identified from IRAC and succeeded in extracting for nearly half of them. From 2MASS archives,  
we have considered only the data with tags of A, B or C (or a S/N of $\geq$ 5) in all the JHK bands.

\begin{table*}
\centering
\caption{{\it Spitzer} IRAC 4-channel photometry (in mag) of the Class 0/I YSOs identified in M8 (see text for details); the
numbers in the last column refer to: 1. 2MASS; 2. Arias et al. (2006); 3. Arias et al. (2007)}

\label{tab1}
\begin{tabular}{lcccccc||ccccccccc}
\hline
Object            &        RA [2000]      &	Dec [2000]    &     Ch1    &	Ch2	&     Ch3    &     Ch4    & 
$\alpha_{IRAC}$     & 
  H  &       J-H     &     H-K       & NIR Ref. \\     
\hline          			  	 	      	 	   	 	 	     	 	  	      
      1 	  &  	   18:02:25.56    &	-24:32:41.7   &    12.63   &	11.88	&    10.87   &	   9.76   &	0.51   &          &		  &		  &		  \\
      2 	  &  	   18:02:29.28    &	-24:08:15.8   &     5.79   &	 5.12	&     4.41   &	   3.45   &    -0.15   &  9.46    &	   3.16   &	  1.96    &  1  	  \\
      3 	  &  	   18:02:42.07    &	-24:19:13.2   &    12.19   &	10.63	&     9.54   &	   8.62   &	1.20   &          &		  &		  &		  \\
      4 	  &  	   18:02:42.97    &	-24:23:25.5   &    11.28   &	10.13	&     9.28   &	   8.33   &	0.51   &          &		  &		  &		  \\
      5 	  &  	   18:02:49.71    &	-24:22:26.6   &    11.36   &	 9.90	&     9.13   &	   8.82   &    -0.01   &          &		  &		  &		  \\
      6 	  &  	   18:02:52.85    &	-24:20:46.6   &    11.52   &	10.27	&     9.54   &	   8.85   &	0.15   &          &		  &		  &		  \\
      7 	  &  	   18:02:53.21    &	-24:20:17.3   &     8.91   &	 8.31	&     7.82   &	   5.94   &	0.49   &  9.79    &	   0.16   &	  0.19    &  1  	  \\
      8 	  &  	   18:03:06.89    &	-24:20:57.6   &    11.08   &	10.29	&     9.52   &	   8.24   &	0.40   &          &		  &		  &		  \\
      9 	  &  	   18:03:06.99    &	-24:21:20.1   &     8.78   &	 7.87	&     6.86   &	   5.94   &	0.46   &          &		  &		  &		  \\
     10 	  &  	   18:03:11.64    &	-24:11:57.0   &     6.63   &	 5.57	&     4.28   &	   3.41   &	0.94   &          &		  &		  &		  \\
     11 	  &  	   18:03:12.28    &	-24:33:30.5   &     8.12   &	 7.13	&     6.16   &	   5.88   &    -0.22   &          &		  &		  &		  \\
     12 	  &  	   18:03:25.59    &	-24:21:08.3   &    10.88   &	10.21	&     9.42   &	   8.55   &    -0.13   &          &		  &		  &		  \\
     13 	  &  	   18:03:27.38    &	-24:21:04.0   &    11.81   &	 9.98	&     9.08   &	   8.63   &	0.69   &          &		  &		  &		  \\
     14 	  &  	   18:03:29.22    &	-24:21:49.9   &    10.34   &	 9.89	&     9.05   &	   7.80   &	0.13   &  13.27   &	    1.44  &	   1.08   &  1  	  \\
     15 	  &  	   18:03:30.84    &	-24:20:03.1   &    11.73   &	 9.94	&     8.63   &	   7.36   &	2.12   &          &		  &		  &		  \\
     16 	  &  	   18:03:35.58    &	-24:22:04.6   &    11.86   &	10.75	&     9.51   &	   7.37   &	2.31   &          &		  &		  &		  \\
     17 	  &  	   18:03:36.21    &	-24:18:13.7   &     9.87   &	 8.67	&     7.91   &	   7.11   &	0.27   &          &		  &		  &		  \\
     18 	  &  	   18:03:36.28    &	-24:17:51.1   &     8.75   &	 7.66	&     6.68   &	   5.53   &	0.85   &          &		  &		  &		  \\
     19 	  &  	   18:03:37.05    &	-24:22:31.6   &    10.34   &	 9.36	&     8.25   &	   7.06   &	0.96   &  13.87   &	    1.13  &	   0.98   &  1,2	  \\
     20 	  &  	   18:03:37.41    &	-24:13:56.9   &     4.64   &	 4.04	&     3.09   &	   2.43   &    -0.21   &  6.83    &	    1.95  &	   1.28   &  1  	  \\
     21 	  &  	   18:03:37.74    &	-24:25:30.3   &    11.24   &	10.18	&     9.47   &	   8.17   &	0.61   &          &		  &		  &		  \\
     22 	  &  	   18:03:38.31    &	-24:33:59.8   &    11.71   &	10.59	&     9.67   &	   8.56   &	0.75   &          &		  &		  &		  \\
     23 	  &  	   18:03:38.65    &	-24:22:24.3   &     9.00   &	 7.90	&     6.96   &	   5.22   &	1.45   & 12.52    &	    1.01  &	   0.92   &  1,2,3	  \\
     24(KS 1)	  &  	   18:03:40.37    &	-24:22:38.6   &     6.19   &	 5.29	&     4.46   &	   2.51   &	1.34   & 10.61    &	    1.63  &	   1.08	  &  2  	  \\
     25(Her 36)   &  	   18:03:40.43    &	-24:22:43.6   &     5.42   &	 4.72	&     3.86   &	   2.27   &	0.79   & 7.45     &	    0.49  &	   0.54   &  1,2	  \\
     26(KS 4)	  &  	   18:03:41.50    &	-24:22:44.3   &     9.36   &	 8.58	&     ---    &	    ---   &	 ---   & 12.99    &         2.17  &        1.46   &  2  	  \\
     27 	  &  	   18:03:44.95    &	-24:16:08.7   &    11.13   &	10.66	&     9.88   &	   8.58   &	0.13   &          &		  &		  &		  \\
     28 	  &  	   18:03:45.16    &	-24:23:25.0   &     8.88   &	 8.20	&     7.54   &	   6.24   &	0.18   & 11.36    &         1.30  &        0.92   & 2,3 	  \\
     29 	  &  	   18:03:47.05    &	-24:25:37.4   &     8.45   &	 7.86	&     6.95   &	   6.08   &    -0.05   & 11.81    &	    1.74  &	   1.37   &  1  	  \\
     30 	  &  	   18:03:47.37    &	-24:18:44.4   &     8.73   &	 8.16	&     7.29   &	   6.33   &    -0.03   &          &		  &		  &		  \\
     31 	  &  	   18:03:47.47    &	-24:25:34.5   &     8.73   &	 7.87	&     6.99   &	   6.13   &	0.16   &          &		  &		  &		  \\
     32 	  &  	   18:03:48.44    &	-24:26:32.0   &     9.77   &	 8.64	&     7.75   &	   7.29   &	0.00   &          &		  &		  &		  \\
     33 	  &  	   18:03:48.47    &	-24:25:58.6   &    10.47   &	 9.09	&     8.40   &	   7.66   &	0.28   &          &		  &		  &		  \\
     34 	  &  	   18:03:50.26    &	-24:22:23.4   &    10.03   &	 9.54	&     8.95   &	   7.80   &    -0.28   & 11.96    &	    0.85  &	   0.56   &  1  	  \\
     35 	  &  	   18:03:54.23    &	-24:25:33.3   &    10.15   &	 9.40	&     8.28   &	   6.88   &	0.98   & 13.27    &	    1.72  &	   1.04   &  1  	  \\
     36 	  &  	   18:04:05.11    &	-24:16:42.3   &    10.80   &	10.04	&     9.56   &	   8.43   &    -0.19   & 12.59    &	    0.83  &	   0.32   &  1  	  \\
     37 	  &  	   18:04:08.81    &	-24:27:27.5   &    10.36   &	 9.54	&     8.90   &	   8.11   &    -0.28   &          &		  &		  &		  \\
     38 	  &  	   18:04:10.55    &	-24:26:56.1   &     9.00   &	 8.32	&     7.84   &	   6.59   &    -0.13   & 11.10    &	    0.99  &	   0.62   &  1,3	  \\
     39 	  &  	   18:04:11.03    &	-24:27:20.5   &    11.66   &	 9.48	&     8.11   &	   7.10   &	2.29   &          &		  &		  &		  \\
     40 	  &  	   18:04:11.07    &	-24:21:31.5   &     6.56   &	 5.72	&     4.83   &	   3.96   &	0.17   &          &		  &		  &		  \\
     41 	  &  	   18:04:11.08    &	-24:26:54.2   &    10.07   &	 8.46	&     7.35   &	   6.39   &	1.33   &          &		  &		  &		  \\
     42 	  &  	   18:04:17.06    &	-24:28:10.9   &    11.32   &	10.62	&     9.87   &	   8.72   &	0.15   & 13.67    &	    1.13  &	   0.65   &  1  	  \\
     43 	  &  	   18:04:20.09    &	-24:29:14.7   &     5.56   &	 4.33	&     2.95   &	   2.77   &	0.46   &          &		  &		  &		  \\
     44 	  &  	   18:04:20.74    &	-24:28:22.1   &    10.06   &	 9.35	&     8.57   &	   7.50   &	0.11   & 13.03    &	    1.87  &	   1.20   &  1  	  \\
     45 	  &  	   18:04:20.95    &	-24:21:07.9   &    11.40   &	10.78	&    10.44   &	   9.01   &    -0.19   & 13.10    &	    1.06  &	   0.56   &  1  	  \\
     46 	  &  	   18:04:21.24    &	-24:28:03.5   &     9.86   &	 9.45	&     8.60   &	   7.39   &	0.07   & 12.49    &	    1.31  &	   0.88   &  1,3	  \\
     47 	  &  	   18:04:21.63    &	-24:11:26.3   &     9.33   &	 8.68	&     7.92   &	   7.08   &    -0.23   & 12.61    &	    2.35  &	   1.51   &  1  	  \\
     48 	  &  	   18:04:21.70    &	-24:21:14.6   &    11.47   &	10.98	&    10.29   &	   9.05   &    -0.04   &          &		  &		  &		  \\
     49 	  &  	   18:04:23.98    &	-24:21:27.0   &     9.19   &	 8.79	&     8.30   &	   6.58   &	0.12   &  9.25    &	    0.07  &	   0.03   &  1  	  \\
     50 	  &  	   18:04:24.31    &	-24:20:59.7   &     9.45   &	 9.17	&     8.49   &	   6.82   &	0.20   &  9.83    &	    0.04  &	   0.01   &  1  	  \\
     51 	  &  	   18:04:26.52    &	-24:29:00.3   &     9.72   &	 9.30	&     8.59   &	   7.36   &    -0.09   & 11.82    &	    1.04  &	   0.50   &  1  	  \\
     52 	  &  	   18:04:28.09    &	-24:22:41.8   &    11.25   &	10.82	&    10.10   &	   8.89   &    -0.10   &          &		  &		  &		  \\
     53 	  &  	   18:04:28.90    &	-24:14:02.6   &     8.46   &	 8.05	&     7.30   &	   6.06   &    -0.04   & 10.19    &	    0.42  &	   0.32   &  1  	  \\
     54 	  &  	   18:04:30.74    &	-24:28:45.6   &     6.75   &	 6.02	&     5.34   &	   4.27   &    -0.01   & 10.52    &	    1.99  &	   1.49   &  1  	  \\
     55 	  &  	   18:04:35.78    &	-24:28:35.6   &     9.03   &	 8.44	&     7.71   &	   6.74   &    -0.19   &          &		  &		  &		  \\
     56 	  &  	   18:04:40.67    &	-24:12:16.9   &    11.06   &	10.66	&     9.96   &	   8.78   &    -0.19   & 13.01    &	    1.06  &	   0.52   &  1  	  \\
     57 	  &  	   18:04:44.19    &	-24:15:25.1   &    10.67   &	10.25	&     9.57   &	   8.32   &    -0.11   & 12.01    &	    0.91  &	   0.37   &  1  	  \\
     58 	  &  	   18:04:47.09    &	-24:27:55.4   &     9.05   &	 8.08	&     7.36   &	   6.51   &	0.04   &          &		  &		  &		  \\
     59 	  &  	   18:04:48.42    &	-24:27:53.8   &    10.99   &	10.66	&     9.92   &	   8.64   &    -0.09   &          &		  &		  &		  \\
     60 	  &  	   18:04:50.37    &	-24:14:25.8   &     5.02   &	 4.27	&     3.29   &	   2.31   &	0.32   &          &		  &		  &		  \\
\hline          
\end{tabular}
\end{table*}

\begin{table*}
\centering
\contcaption{}
\label{tab1}
\begin{tabular}{lcccccc||cccccccccc}
\hline
Object            &         RA [2000]     &	 Dec [2000]   &      Ch1   &	 Ch2	&     Ch3    &	   Ch4    &    $\alpha_{IRAC}$  &  
 H      &   J-H	&     H-K	& NIR Ref.   \\
\hline                     		   		         	    	 	 	      	 	   	      						 
     61  	  &  	   18:04:50.62    &	-24:25:42.2   &     8.73   &	 8.11	&     7.67   &	   5.87   &	0.36  &      10.16   &   0.29	&     0.43	& 3		  \\
     62  	  &  	   18:04:51.13    &	-24:26:33.7   &     9.48   &	 9.03	&     8.52   &	   6.93   &	0.05  &      12.68   &   1.95	&     1.13	& 1		  \\
     63  	  &  	   18:04:56.77    &	-24:27:16.4   &    10.30   &	 9.24	&     8.44   &	   7.56   &	0.27  & 	     &  	&		&		  \\
     64  	  &  	   18:04:58.82    &	-24:26:24.1   &     9.54   &	 8.47	&     7.82   &	   6.89   &	0.13  & 	     &  	&		&		  \\
     65  	  &  	   18:05:00.47    &	-24:13:26.8   &     9.51   &	 8.88	&     8.25   &	   6.94   &	0.08  &      10.89   &   0.62	&     0.37	& 1		  \\
\hline 
\end{tabular}
\end{table*}

\begin{table*}
\centering
\caption{{\it Spitzer} IRAC 4-channel photometry (in mag) of the Class II YSOs identified in M8 (see text for details); the
numbers in the last column refer to: 1. 2MASS; 2. Arias et al. (2006); 3. Arias et al. (2007)}

\label{tab2}
\begin{tabular}{lcccccc||ccccccccc}
\hline
Object        &    RA [2000]    &     Dec [2000]   &     Ch1    &    Ch2    &    Ch3      &    Ch4    &   $\alpha_{IRAC}$    &    H      & 
 J-H    &     H-K        &   NIR Ref.  \\	   
\hline          											    	  	
           1  &  18:02:23.78	&    -24:08:49.4   &	10.76	&    10.41  &	  10.04   &	9.26  &     -1.13  &	  12.29  &	0.94   &     0.42   &  1  \\ 
           2  &  18:02:24.95	&    -24:36:08.4   &	 7.92	&     7.57  &	   7.12   &	6.86  &     -1.59  &	   9.59  &	2.31   &     1.02   &  1  \\ 
           3  &  18:02:25.98	&    -24:27:31.8   &	 6.83	&     6.25  &	   5.59   &	4.74  &     -0.42  &	  11.46  &	3.65   &     2.14   &  1  \\ 
           4  &  18:02:27.52	&    -24:22:52.9   &	 8.12	&     7.53  &	   6.69   &	6.49  &     -0.88  &	         &	       &	    &	  \\ 
           5  &  18:02:27.65	&    -24:11:38.0   &	11.53	&    11.20  &	  10.43   &	9.70  &     -0.66  &	         &	       &	    &	  \\ 
           6  &  18:02:29.49	&    -24:24:53.7   &	10.58	&    10.05  &	   9.81   &	8.97  &     -1.07  &	  12.71  &	1.08   &     0.69   &  1  \\ 
           7  &  18:02:36.33	&    -24:21:07.9   &	 8.10	&     7.83  &	   7.23   &	6.94  &     -1.43  &	  11.15  &	3.46   &     1.78   &  1  \\ 
           8  &  18:02:37.34	&    -24:16:24.4   &	10.40	&     9.98  &	   9.55   &	8.86  &     -1.08  &	  12.47  &	0.92   &     0.56   &  1  \\ 
           9  &  18:02:41.45	&    -24:10:35.3   &	 8.26	&     7.91  &	   7.36   &	7.20  &     -1.56  &	  10.85  &	3.13   &     1.48   &  1  \\ 
          10  &  18:02:41.65	&    -24:33:55.0   &	 7.92	&     7.55  &	   6.94   &	6.11  &     -0.73  &	   9.60  &	2.07   &     0.93   &  1  \\ 
          11  &  18:02:42.67	&    -24:17:18.9   &	 6.34	&     6.06  &	   5.42   &	4.88  &     -1.09  &	         &	       &	    &	  \\ 
          12  &  18:02:43.91	&    -24:15:23.1   &	10.71	&    10.26  &	   9.85   &	9.03  &     -0.93  &	  12.55  &	0.77   &     0.56   &  1  \\ 
          13  &  18:02:45.65	&    -24:13:49.7   &	 8.47	&     7.92  &	   7.53   &	7.24  &     -1.44  &	  10.85  &	2.89   &     1.47   &  1  \\ 
          14  &  18:02:48.85	&    -24:21:08.8   &	 9.56	&     9.24  &	   8.87   &	7.88  &     -0.92  &	  12.34  &	0.84   &     0.62   &  1  \\ 
          15  &  18:02:49.10	&    -24:28:15.8   &	 7.32	&     6.91  &	   6.47   &	6.13  &     -1.46  &	  10.59  &	2.99   &     1.62   &  1  \\ 
          16  &  18:02:49.15	&    -24:08:52.3   &	 6.50	&     6.04  &	   5.73   &	4.96  &     -1.12  &	   9.75  &	2.88   &     1.59   &  1  \\ 
          17  &  18:02:49.33	&    -24:11:19.0   &	 5.80	&     5.34  &	   5.00   &	4.63  &     -1.51  &	   9.30  &	2.55   &     1.47   &  1  \\ 
          18  &  18:02:50.81	&    -24:17:56.6   &	 9.95	&     9.29  &	   9.03   &	8.29  &     -1.02  &	  12.50  &	1.43   &     0.95   &  1  \\ 
          19  &  18:02:50.95	&    -24:22:20.1   &	10.65	&    10.21  &	   9.83   &	8.82  &     -0.76  &	         &	       &	    &	  \\ 
          20  &  18:02:51.10	&    -24:19:23.4   &	 9.59	&     9.20  &	   8.84   &	8.02  &     -1.05  &	  12.64  &	1.32   &     0.95   &  1  \\ 
          21  &  18:02:51.12	&    -24:16:57.2   &	 5.29	&     4.20  &	   3.38   &	3.18  &     -0.43  &	         &	       &	    &	  \\ 
          22  &  18:02:51.15	&    -24:18:07.9   &	10.45	&     9.78  &	   9.50   &	8.74  &     -0.96  &	  12.95  &	0.96   &     0.71   &  1  \\ 
          23  &  18:02:51.39	&    -24:17:13.3   &	 9.48	&     9.15  &	   8.59   &	8.26  &     -1.38  &	         &	       &	    &	  \\ 
          24  &  18:02:52.47	&    -24:18:44.7   &	 8.56	&     8.14  &	   7.64   &	6.78  &     -0.79  &	  10.72  &	0.90   &     0.75   &  1  \\ 
          25  &  18:02:53.83	&    -24:20:19.8   &	 9.22	&     8.68  &	   8.60   &	7.92  &     -1.45  &	         &	       &	    &	  \\ 
          26  &  18:02:53.96	&    -24:20:11.3   &	 9.97	&     9.51  &	   9.16   &	8.41  &     -1.09  &	  12.61  &	1.29   &     1.00   &  1  \\ 
          27  &  18:02:54.30	&    -24:20:56.5   &	 8.85	&     8.35  &	   8.01   &	7.20  &     -0.99  &	  11.23  &	0.92   &     0.77   &  1  \\ 
          28  &  18:02:54.72	&    -24:19:56.9   &	 9.57	&     9.14  &	   8.75   &	8.47  &     -1.57  &	         &	       &	    &	  \\ 
          29  &  18:02:56.79	&    -24:23:34.8   &	10.17	&     9.89  &	   9.62   &	8.98  &     -1.49  &	  12.32  &	2.42   &     1.10   &  1  \\ 
          30  &  18:02:56.88	&    -24:35:40.7   &	 8.08	&     7.79  &	   7.28   &	7.04  &     -1.59  &	  10.20  &	2.84   &     1.29   &  1  \\ 
          31  &  18:03:01.50	&    -24:27:47.1   &	10.53	&    10.13  &	   9.47   &	8.48  &     -0.45  &	  12.93  &	1.70   &     0.90   &  1  \\ 
          32  &  18:03:05.14	&    -24:31:54.2   &	 5.57	&     5.16  &	   4.61   &	3.81  &     -0.80  &	         &	       &	    &	  \\ 
          33  &  18:03:06.56	&    -24:32:13.1   &	 5.60	&     4.96  &	   4.44   &	3.56  &     -0.53  &	   8.46  &	2.91   &     1.63   &  1  \\ 
          34  &  18:03:09.96	&    -24:33:50.4   &	 8.10	&     7.50  &	   6.97   &	6.67  &     -1.19  &	  10.79  &	2.63   &     1.38   &  1  \\ 
          35  &  18:03:10.31	&    -24:26:59.5   &	 6.49	&     5.89  &	   5.43   &	4.59  &     -0.69  &	  10.50  &	2.93   &     1.92   &  1  \\ 
          36  &  18:03:17.69	&    -24:20:53.2   &	10.74	&    10.25  &	   9.77   &	8.93  &     -0.77  &	  12.20  &	1.08   &     0.45   &  1  \\ 
          37  &  18:03:18.30	&    -24:25:58.7   &	 8.19	&     7.44  &	   6.70   &	6.20  &     -0.53  &	  10.30  &	2.58   &     1.45   &  1  \\ 
          38  &  18:03:20.03	&    -24:20:20.4   &	10.45	&    10.21  &	   9.81   &	9.05  &     -1.21  &	  12.45  &	2.12   &     0.99   &  1  \\ 
          39  &  18:03:20.53	&    -24:30:29.8   &	 5.27	&     5.09  &	   4.69   &	4.18  &     -1.54  &	   8.31  &	2.35   &     1.44   &  1  \\ 
          40  &  18:03:23.06	&    -24:23:41.8   &	 7.03	&     6.61  &	   6.22   &	5.82  &     -1.45  &	  11.43  &	2.97   &     1.82   &  1  \\ 
          41  &  18:03:23.09	&    -24:21:33.1   &	 9.80	&     9.53  &	   8.88   &	8.65  &     -1.42  &	         &	       &	    &	  \\ 
          42  &  18:03:24.06	&    -24:21:23.3   &	 8.58	&     7.94  &	   7.17   &	6.95  &     -0.91  &	         &	       &	    &	  \\ 
          43  &  18:03:25.06	&    -24:19:02.2   &	 8.07	&     7.72  &	   7.07   &	6.93  &     -1.44  &	  10.34  &	3.06   &     1.45   &  1  \\ 
          44  &  18:03:25.12	&    -24:21:28.4   &	11.06	&    10.06  &	   9.49   &	8.85  &     -0.36  &	         &	       &	    &	  \\ 
          45  &  18:03:26.71	&    -24:22:11.3   &	 8.91	&     8.68  &	   8.27   &	7.22  &     -0.88  &	         &	       &	    &	  \\ 
          46  &  18:03:34.15	&    -24:24:59.7   &	 8.17	&     7.83  &	   7.19   &	7.07  &     -1.50  &	  10.50  &	2.96   &     1.44   &  1  \\ 
          47  &  18:03:34.74	&    -24:18:53.7   &	 6.44	&     6.20  &	   5.68   &	5.32  &     -1.50  &	  10.57  &	3.94   &     2.21   &  1  \\ 
          48  &  18:03:35.86	&    -24:09:11.9   &	 9.61	&     8.85  &	   8.82   &	7.90  &     -1.04  &	  11.63  &	0.91   &     0.64   &  1  \\ 
          49  &  18:03:36.29	&    -24:09:48.2   &	 6.11	&     5.74  &	   5.31   &	4.38  &     -0.85  &	  10.88  &	2.82   &     1.78   &  1  \\ 
          50  &  18:03:36.68	&    -24:10:32.6   &	 8.59	&     8.39  &	   8.00   &	7.52  &     -1.58  &	  10.50  &	2.04   &     0.86   &  1  \\ 
\hline          
\end{tabular}
\end{table*}

\begin{table*}
\centering
\contcaption{}
\label{tab2}
\begin{tabular}{lcccccc||ccccccccc}
\hline
Object          &   RA [2000]    &   Dec [2000]      &      Ch1    &     Ch2      &     Ch3    &      Ch4    &    $\alpha_{IRAC}$    &    H
     &   J-H    &     H-K         &   NIR Ref.  \\
\hline                                                                                                        		      
          51    & 18:03:36.84	 &     -24:24:15.1   &     10.23   &	  9.71    &	9.39   &      8.20   &     -0.58   &	12.26  &      0.89  &	   0.37  &  1	  \\  
          52    & 18:03:37.32	 &     -24:22:46.9   &	    9.01   &	  8.44    &	7.99   &      6.88   &	   -0.44   &	11.47  &      1.34  &	   0.66	 &  2	  \\
          53    & 18:03:38.49	 &     -24:22:31.7   &	    8.81   &	  8.11    &	7.57   &      6.63   &	   -0.38   &	10.80  &      1.22  &	   0.80  &  1,2   \\
          54    & 18:03:38.81	 &     -24:08:58.8   &	    5.43   &	  5.08    &	4.59   &      3.86   &	   -1.02   &	 8.58  &      2.92  &	   1.55  &  1	  \\
          55    & 18:03:39.05	 &     -24:28:10.8   &	    8.16   &	  7.74    &	7.09   &      6.81   &	   -1.23   &	 9.97  &      2.41  &	   1.11  &  1	  \\
          56    & 18:03:39.38	 &     -24:25:24.2   &	    9.54   &	  9.06    &	8.54   &      7.38   &	   -0.37   &	11.85  &      1.14  &	   0.69  &  1	  \\
          57    & 18:03:40.23	 &     -24:22:03.8   &	    9.54   &	  9.15    &	8.74   &      7.84   &	   -0.90   &	11.66  &      1.31  &	   0.63  &  1	  \\
          58    & 18:03:40.24	 &     -24:29:08.2   &	    5.05   &	  4.51    &	3.84   &      2.93   &	   -0.38   &	 8.93  &      2.76  &	   1.92  &  1	  \\
          59    & 18:03:40.33	 &     -24:25:05.2   &	   10.39   &	  9.78    &	9.54   &      8.58   &	   -0.86   &	12.25  &      1.04  &	   0.54  &  1	  \\
          60    & 18:03:40.74	 &     -24:23:16.3   &	    8.67   &	  8.29    &	7.83   &      6.87   &	   -0.77   &	11.24  &      1.10  &	   0.80  &  1,2,3 \\
          61    & 18:03:41.05	 &     -24:25:45.6   &	    8.54   &	  8.21    &	7.73   &      7.23   &	   -1.30   &	       &	    &		 &	  \\
          62    & 18:03:42.26	 &     -24:23:22.4   &	    9.68   &	  9.10    &	8.70   &      7.76   &	   -0.69   &	12.48  &      1.40  &	   0.94  &  1,2   \\
          63    & 18:03:42.89	 &     -24:16:26.4   &	    8.17   &	  7.79    &	7.21   &      7.03   &	   -1.47   &	 9.83  &      2.26  &	   1.07  &  1	  \\
          64    & 18:03:43.09	 &     -24:21:29.6   &	    8.49   &	  8.29    &	7.87   &      7.08   &	   -1.20   &	10.84  &      2.75  &	   1.25  &  1	  \\
          65    & 18:03:43.29	 &     -24:28:07.1   &	    8.14   &	  7.90    &	7.19   &      7.09   &	   -1.52   &	10.13  &      2.62  &	   1.29  &  1	  \\
          66    & 18:03:45.07	 &     -24:22:05.6   &	    8.61   &	  8.45    &	8.15   &      6.37   &	   -0.30   &	 9.13  &      0.11  &	   0.10  &  1,2   \\
          67    & 18:03:47.93	 &     -24:18:02.1   &	    9.05   &	  8.45    &	8.00   &      7.37   &	   -0.94   &	12.65  &      1.94  &	   1.43  &  1	  \\
          68    & 18:03:49.38	 &     -24:26:14.6   &	    8.47   &	  8.04    &	7.42   &      7.27   &	   -1.40   &	       &	    &		 &	  \\
          69    & 18:03:49.61	 &     -24:22:09.3   &	   10.85   &	 10.36    &	9.99   &      8.85   &	   -0.59   &	12.65  &      0.99  &	   0.42  &  1	  \\
          70    & 18:03:50.73	 &     -24:20:13.3   &	    9.81   &	  9.57    &	9.31   &      8.44   &	   -1.29   &	       &	    &		 &	  \\
          71    & 18:03:50.79	 &     -24:21:10.9   &	    6.46   &	  6.14    &	5.61   &      4.60   &	   -0.68   &	 9.21  &      0.86  &	   0.85  &  1,3   \\
          72    & 18:03:51.65	 &     -24:28:26.7   &	    9.51   &	  9.21    &	8.91   &      8.12   &	   -1.25   &	11.58  &      0.80  &	   0.64  &  1	  \\
          73    & 18:03:53.05	 &     -24:14:51.6   &	    8.17   &	  7.59    &	7.17   &      6.89   &	   -1.38   &	       &	    &		 &	  \\
          74    & 18:03:57.82	 &     -24:20:51.3   &	   10.21   &	  9.67    &	9.43   &      8.52   &	   -0.97   &	       &	    &		 &	  \\
          75    & 18:03:57.84	 &     -24:25:34.9   &	    6.09   &	  5.54    &	5.09   &      4.25   &	   -0.76   &	 8.84  &      0.94  &	   0.83  &  1	  \\
          76    & 18:03:58.29	 &     -24:16:49.3   &	    9.95   &	  9.26    &	8.95   &      8.10   &	   -0.81   &	12.24  &      1.02  &	   0.71  &  1,3   \\
          77    & 18:03:58.54	 &     -24:24:58.8   &	    9.66   &	  9.31    &	8.90   &      8.18   &	   -1.14   &	11.24  &      0.87  &	   0.41  &  1	  \\
          78    & 18:03:59.27	 &     -24:23:08.2   &	    9.01   &	  8.79    &	8.59   &      7.76   &	   -1.44   &	11.16  &      1.24  &	   0.93  &  1	  \\
          79    & 18:04:01.13	 &     -24:22:35.6   &	   10.49   &	 10.30    &	9.93   &      8.96   &	   -1.06   &	12.33  &      2.19  &	   0.99  &  1	  \\
          80    & 18:04:03.10	 &     -24:25:19.5   &	    9.53   &	  9.17    &	8.70   &      7.60   &	   -0.63   &	11.72  &      0.66  &	   0.58  &  1	  \\
          81    & 18:04:04.65	 &     -24:08:49.8   &	    8.03   &	  7.69    &	7.13   &      6.92   &	   -1.51   &	 9.89  &      2.31  &	   1.12  &  1	  \\
          82    & 18:04:07.89	 &     -24:26:06.3   &	   10.69   &	 10.27    &	9.69   &      8.55   &	   -0.37   &	       &	    &		 &	  \\
          83    & 18:04:08.11	 &     -24:20:55.6   &	   10.90   &	 10.39    &    10.04   &      9.04   &	   -0.75   &	13.03  &      1.00  &	   0.47  &  1	  \\
          84    & 18:04:08.47	 &     -24:20:49.5   &	   10.76   &	 10.44    &    10.08   &      8.93   &	   -0.76   &	12.39  &      0.93  &	   0.38  &  1	  \\
          85    & 18:04:09.94	 &     -24:25:32.7   &	    9.72   &	  9.27    &	9.02   &      8.20   &	   -1.15   &	11.15  &      0.94  &	   0.38  &  1	  \\
          86    & 18:04:10.19	 &     -24:25:49.6   &	    9.09   &	  8.62    &	8.20   &      7.27   &	   -0.78   &	11.80  &      1.17  &	   0.69  &  1	  \\
          87    & 18:04:11.57	 &     -24:28:42.1   &	   12.10   &	 11.85    &    11.11   &     10.95   &	   -1.40   &	       &	    &		 &	  \\
          88    & 18:04:11.99	 &     -24:26:28.1   &	   10.18   &	  9.57    &	9.07   &      8.18   &	   -0.57   &	11.81  &      0.96  &	   0.43  &  1	  \\
          89    & 18:04:12.48	 &     -24:11:51.6   &	   10.83   &	 10.10    &	9.70   &      8.67   &	   -0.44   &	12.83  &      1.00  &	   0.73  &  1	  \\
          90    & 18:04:12.52	 &     -24:35:47.1   &	    8.15   &	  7.83    &	7.43   &      7.07   &	   -1.58   &	 9.39  &      2.02  &	   0.81  &  1	  \\
          91    & 18:04:13.10	 &     -24:26:13.3   &	   10.88   &	 10.43    &	9.90   &      8.80   &	   -0.46   &	12.38  &      0.99  &	   0.43  &  1	  \\
          92    & 18:04:15.75	 &     -24:19:01.7   &	   10.07   &	  9.80    &	9.64   &      8.80   &	   -1.43   &	11.71  &      0.80  &	   0.41  &  1,3   \\
          93    & 18:04:15.77	 &     -24:25:15.8   &	   10.71   &	 10.17    &	9.79   &      9.09   &	   -1.02   &	       &	    &		 &	  \\
          94    & 18:04:15.91	 &     -24:18:46.2   &	   10.05   &	  9.35    &	9.12   &      8.48   &	   -1.13   &	12.44  &      1.11  &	   0.81  &  1,3   \\
          95    & 18:04:16.06	 &     -24:27:57.4   &	   10.05   &	  9.70    &	9.34   &      8.79   &	   -1.40   &	       &	    &		 &	  \\
          96    & 18:04:16.14	 &     -24:19:52.5   &	    9.41   &	  8.86    &	8.44   &      7.45   &	   -0.63   &	11.45  &      0.92  &	   0.62  &  1	  \\
          97    & 18:04:16.41	 &     -24:24:38.8   &	    9.67   &	  9.17    &	8.79   &      7.74   &	   -0.67   &	11.88  &      1.10  &	   0.78  &  1,3   \\
          98    & 18:04:16.93	 &     -24:24:14.9   &	    9.16   &	  8.75    &	8.38   &      7.84   &	   -1.33   &	10.96  &      0.70  &	   0.38  &  1	  \\
          99    & 18:04:17.42	 &     -24:19:09.8   &	    8.83   &	  8.42    &	8.00   &      7.53   &	   -1.34   &	11.37  &      1.57  &	   0.98  &  1,3   \\
         100    & 18:04:17.89	 &     -24:17:46.8   &	   10.39   &	  9.84    &	9.60   &      8.44   &	   -0.69   &	12.02  &      0.94  &	   0.57  &  1	  \\
         101    & 18:04:19.09	 &     -24:27:58.8   &	   10.28   &	  9.70    &	9.17   &      8.18   &	   -0.46   &	12.43  &      1.20  &	   0.73  &  1	  \\
         102    & 18:04:19.31	 &     -24:22:54.9   &	    9.95   &	  9.60    &	9.33   &      8.70   &	   -1.42   &	11.50  &      0.94  &	   0.47  &  1,3   \\
         103    & 18:04:19.58	 &     -24:24:04.7   &	    9.00   &	  8.77    &	8.37   &      7.81   &	   -1.44   &	11.79  &      1.48  &	   1.16  &  1	  \\
         104    & 18:04:19.87	 &     -24:28:23.7   &	    9.45   &	  8.86    &	8.46   &      7.73   &	   -0.91   &	       &	    &		 &	  \\
         105    & 18:04:20.07	 &     -24:22:48.2   &	   10.58   &	 10.26    &	9.66   &      8.51   &	   -0.44   &	12.09  &      0.89  &	   0.83	 &  3	  \\
         106    & 18:04:20.26	 &     -24:20:24.8   &	   10.14   &	  9.79    &	9.41   &      8.65   &	   -1.14   &	12.32  &      1.04  &	   0.73  &  1	  \\
         107    & 18:04:20.34	 &     -24:24:34.6   &	   10.75   &	  9.93    &	9.51   &      9.07   &	   -0.96   &	       &	    &		 &	  \\
         108    & 18:04:20.52	 &     -24:23:04.1   &	    9.96   &	  9.48    &	8.97   &      8.12   &	   -0.73   &	       &	    &		 &	  \\
         109    & 18:04:20.61	 &     -24:23:01.1   &	   10.24   &	  9.89    &	9.45   &      8.47   &	   -0.81   &	       &	    &		 &	  \\
         110    & 18:04:20.82	 &     -24:23:22.5   &	   10.33   &	  9.92    &	9.68   &      8.88   &	   -1.22   &	12.46  &      1.05  &	   0.83  &  1	  \\
\hline          												    
\end{tabular}													    
\end{table*}													    
														    
\begin{table*}													    
\centering													    
\contcaption{}													    
\label{tab2}													    
\begin{tabular}{lcccccc||ccccccccc}										    
\hline														    
Object        &    RA [2000]  &       Dec [2000]  &     Ch1    &     Ch2    &     Ch3     &     Ch4   &    $\alpha_{IRAC}$    &    H      &
  J-H    &     H-K        &   NIR Ref.   \\
\hline                                                                                                       	    	  	
         111  &  18:04:21.01  &      -24:13:41.8  &	 5.42  &      5.01  &	   4.51   &	3.69  &     -0.85   &	    8.69    &	  2.32  &	1.39	&  1  \\
         112  &  18:04:21.11  &      -24:23:25.5  &	10.08  &      9.81  &	   9.42   &	8.66  &     -1.20   &	   	    &		&		&     \\
         113  &  18:04:21.12  &      -24:20:47.7  &	10.46  &     10.24  &	   9.84   &	8.79  &     -0.91   &	   	    &		&		&     \\
         114  &  18:04:21.19  &      -24:24:22.4  &	10.39  &      9.96  &	   9.55   &	8.89  &     -1.13   &	   	    &		&		&     \\
         115  &  18:04:21.47  &      -24:23:19.1  &	 9.46  &      9.16  &	   8.62   &	7.94  &     -1.05   &	   	    &		&		&     \\
         116  &  18:04:21.69  &      -24:23:19.7  &	 9.75  &      9.27  &	   8.63   &	7.75  &     -0.51   &	   	    &		&		&     \\
         117  &  18:04:21.82  &      -24:22:15.6  &	10.84  &     10.58  &	  10.09   &	9.03  &     -0.74   &	   13.02    &	  0.91  &	0.52	&  1  \\
         118  &  18:04:21.84  &      -24:16:26.2  &	 9.83  &      9.50  &	   9.27   &	8.23  &     -1.05   &	   11.67    &	  0.89  &	0.49	&  1  \\
         119  &  18:04:21.86  &      -24:11:37.5  &	10.57  &     10.24  &	   9.80   &	8.53  &     -0.51   &	   11.77    &	  0.51  &	0.22	&  1  \\
         120  &  18:04:22.77  &      -24:22:09.7  &	 8.53  &      8.09  &	   7.65   &	6.87  &     -0.95   &	    8.86    &	  0.22  &	0.30	&  1  \\
         121  &  18:04:23.05  &      -24:24:15.2  &	10.12  &      9.88  &	   9.27   &	8.94  &     -1.40   &	   	    &		&		&     \\
         122  &  18:04:23.54  &      -24:22:47.6  &	 9.69  &      9.15  &	   8.76   &	8.46  &     -1.45   &	   	    &		&		&     \\
         123  &  18:04:24.23  &      -24:16:25.2  &	 9.95  &      9.53  &	   9.04   &	8.29  &     -0.94   &	   12.71    &	  1.47  &	1.01	&  1  \\
         124  &  18:04:26.15  &      -24:22:45.1  &	 6.73  &      6.24  &	   5.77   &	5.03  &     -0.90   &	   10.23    &	  1.65  &	1.27	&  1  \\
         125  &  18:04:26.75  &      -24:22:42.0  &	11.17  &     10.85  &	  10.35   &	9.08  &     -0.43   &	   13.08    &	  1.09  &	0.62	&  1  \\
         126  &  18:04:26.84  &      -24:23:23.5  &	 9.07  &      8.71  &	   8.44   &	7.71  &     -1.30   &	   11.30    &	  0.88  &	0.66	&  1  \\
         127  &  18:04:27.09  &      -24:21:06.8  &	10.84  &     10.37  &	  10.16   &	9.07  &     -0.89   &	   12.36    &	  0.95  &	0.51	&  1  \\
         128  &  18:04:27.27  &      -24:20:57.0  &	 9.41  &      9.06  &	   8.82   &	8.23  &     -1.52   &	   12.59    &	  1.47  &	1.03	&  1  \\
         129  &  18:04:27.36  &      -24:14:27.3  &	 9.84  &      9.38  &	   8.74   &	8.05  &     -0.75   &	   12.69    &	  1.31  &	1.09	&  1  \\
         130  &  18:04:27.48  &      -24:07:31.1  &	 8.03  &      7.76  &	   7.19   &	6.85  &     -1.42   &	   	    &		&		&     \\
         131  &  18:04:28.03  &      -24:21:43.2  &	 8.03  &      7.57  &	   7.16   &	6.04  &     -0.59   &	    7.97    &	 -0.04  &	0.07	&  1  \\
         132  &  18:04:28.25  &      -24:25:48.0  &	10.14  &      9.76  &	   9.39   &	8.62  &     -1.11   &	   	    &		&		&     \\
         133  &  18:04:28.66  &      -24:20:20.0  &	 5.77  &      5.47  &	   5.16   &	4.44  &     -1.33   &	    8.19    &	  2.12  &	1.18	&  1  \\
         134  &  18:04:29.67  &      -24:25:19.4  &	 7.09  &      6.90  &	   6.26   &	6.03  &     -1.52   &	   10.36    &	  2.07  &	1.31	&  1  \\
         135  &  18:04:29.93  &      -24:14:29.8  &	10.46  &     10.08  &	   9.76   &	9.05  &     -1.24   &	   12.67    &	  1.13  &	0.63	&  1  \\
         136  &  18:04:30.59  &      -24:26:06.9  &	 8.16  &      7.69  &	   7.25   &	6.85  &     -1.33   &	   10.29    &	  1.94  &	1.00	&  1  \\
         137  &  18:04:31.07  &      -24:29:09.9  &	10.59  &     10.41  &	   9.89   &	9.04  &     -1.00   &	   	    &		&		&     \\
         138  &  18:04:32.35  &      -24:19:28.1  &	10.07  &      9.53  &	   9.07   &	8.02  &     -0.52   &	   12.48    &	  0.96  &	0.59	&  1  \\
         139  &  18:04:32.39  &      -24:27:55.2  &	10.33  &      9.95  &	   9.61   &	8.82  &     -1.13   &	   13.36    &	  1.62  &	1.04	&  1  \\
         140  &  18:04:33.22  &      -24:27:18.0  &	 9.68  &      9.02  &	   8.77   &	8.17  &     -1.18   &	   11.56    &	  1.15  &	0.77	&  1  \\
         141  &  18:04:33.58  &      -24:21:54.8  &	 8.70  &      8.45  &	   7.81   &	6.89  &     -0.70   &	   10.42    &	  0.97  &	0.61	&  1  \\
         142  &  18:04:34.61  &      -24:09:02.0  &	 5.30  &      4.66  &	   4.23   &	3.43  &     -0.75   &	    9.29    &	  2.80  &	1.74	&  1  \\
         143  &  18:04:34.94  &      -24:22:51.9  &	10.70  &     10.27  &	   9.92   &	8.84  &     -0.75   &	   12.99    &	  0.99  &	0.58	&  1  \\
         144  &  18:04:36.50  &      -24:19:13.7  &	10.23  &      9.73  &	   9.38   &	8.96  &     -1.40   &	   	    &		&		&     \\
         145  &  18:04:38.90  &      -24:19:28.2  &	 8.24  &      7.87  &	   7.32   &	7.11  &     -1.49   &	   	    &		&		&     \\
         146  &  18:04:39.31  &      -24:32:24.4  &	 5.64  &      5.25  &	   4.61   &	3.77  &     -0.65   &	   10.00    &	  2.60  &	1.61	&  1  \\
         147  &  18:04:39.46  &      -24:27:09.2  &	10.46  &      9.99  &	   9.57   &	8.65  &     -0.79   &	   12.26    &	  1.10  &	0.53	&  1  \\
         148  &  18:04:39.86  &      -24:23:05.2  &	 9.55  &      9.05  &	   8.53   &	7.57  &     -0.58   &	   12.28    &	  1.22  &	0.83	&  1  \\
         149  &  18:04:40.90  &      -24:17:10.8  &	 9.29  &      9.01  &	   8.71   &	8.15  &     -1.54   &	   11.70    &	  1.05  &	0.70	&  1,3\\
         150  &  18:04:41.27  &      -24:15:44.9  &	 5.14  &      4.58  &	   4.14   &	3.83  &     -1.35   &	    7.66    &	  2.63  &	1.41	&  1  \\
         151  &  18:04:41.63  &      -24:26:31.8  &	 8.11  &      7.49  &	   7.02   &	6.71  &     -1.24   &	   10.05    &	  2.07  &	1.11	&  1  \\
         152  &  18:04:43.53  &      -24:27:38.7  &	 8.40  &      7.76  &	   7.22   &	6.46  &     -0.64   &	   10.78    &	  1.21	&	0.76	&  3  \\
         153  &  18:04:43.65  &      -24:27:59.1  &	10.26  &      9.86  &	   9.63   &	9.00  &     -1.44   &	   13.01    &	  1.47  &	0.94	&  1  \\
         154  &  18:04:44.06  &      -24:19:39.7  &	10.48  &     10.02  &	   9.75   &	8.99  &     -1.18   &	   12.07    &	  0.84  &	0.43	&  1  \\
         155  &  18:04:44.46  &      -24:10:17.5  &	10.15  &      9.85  &	   9.64   &	8.85  &     -1.38   &	   	    &		&		&     \\
         156  &  18:04:46.41  &      -24:26:08.0  &	10.42  &     10.05  &	   9.52   &	8.52  &     -0.65   &	   12.54    &	  1.13  &	0.53	&  1  \\
         157  &  18:04:48.05  &      -24:27:24.6  &	 7.20  &      6.59  &	   6.16   &	5.74  &     -1.18   &	   	    &		&		&     \\
         158  &  18:04:48.56  &      -24:26:40.7  &	 8.89  &      8.46  &	   8.04   &	7.29  &     -1.02   &	   11.02    &	  1.58  &	0.95	&  1,3\\
         159  &  18:04:50.23  &      -24:27:59.4  &	10.40  &     10.09  &	   9.36   &	8.82  &     -0.93   &	   	    &		&		&     \\
         160  &  18:04:51.53  &      -24:24:17.4  &	10.06  &      9.45  &	   9.00   &	8.08  &     -0.61   &	   12.28    &	  1.21  &	0.61	&  1  \\
         161  &  18:04:51.57  &      -24:26:10.9  &	10.72  &     10.02  &	   9.96   &	9.00  &     -1.01   &	   	    &		&		&     \\
         162  &  18:04:51.63  &      -24:25:15.8  &	10.37  &      9.88  &	   9.40   &	8.91  &     -1.16   &	   12.02    &	  0.97  &	0.47	&  1  \\
         163  &  18:04:52.64  &      -24:27:30.8  &	10.72  &     10.42  &	   9.67   &	8.99  &     -0.76   &	   13.28    &	  1.16  &	0.70	&  1  \\
         164  &  18:04:53.48  &      -24:26:08.7  &	 8.45  &      7.91  &	   7.21   &	6.66  &     -0.74   &	   	    &		&		&     \\
         165  &  18:04:54.08  &      -24:26:23.7  &	 7.05  &      6.33  &	   5.96   &	5.45  &     -1.06   &	   	    &		&		&     \\
         166  &  18:04:55.00  &      -24:27:18.1  &	 9.15  &      8.77  &	   8.29   &	7.42  &     -0.84   &	   12.59    &	  2.04  &	1.33	&  1  \\
         167  &  18:04:56.18  &      -24:09:00.9  &	 8.16  &      7.78  &	   7.55   &	6.98  &     -1.52   &	   10.06    &	  2.02  &	0.99	&  1  \\
         168  &  18:04:57.35  &      -24:20:48.4  &	 4.49  &      4.30  &	   3.58   &	3.28  &     -1.33   &	   	    &		&		&     \\
\hline          											     	    
\end{tabular}												     	    
\end{table*}													   
														   
\section{Results and Discussion}
\label{Results}
We divide the results and discussion into two subsections: one in which we discuss the IRAC
photometry and the pre-main-sequence sources or the young stellar objects (YSOs) 
detected and their possible formation scenario; and the 
second in which we describe the ratio maps produced from IRAC images and discuss possible 
interpretations and their implications.

\subsection{IRAC Photometry}
Fig 1 shows the IRAC Ch4 (8$\mu$m) image of $\sim$ 42.5$\times$30.0 arcmin$^{2}$ of M8 region with  
Her 36 situated near the centre. Following earlier workers, M8 may be divided into 
a few distinct regions for convenience (see Fig 1): the Her 36 region comprising 
of the bubble-like structure around Her 36 with the massive stars 9 Sgr and 
HD 164816 forming the eastern/north-eastern bounds; the central ridge that includes filamentary structures 
seen in the NE-SW direction; 
the massive star HD 164906 and the cluster NGC 6530; 
the east-west ridge region consisting of the finger-like filamentary 
structures starting from far-east to the south of Her 36; and the compact young cluster region M8E. 

Using the [3.6]-[4.5] vs [5.8]-[8.0] colour-colour diagrams, \citet{Allen04} and \citet{Megeath04} 
formulated division criteria for various pre-main-sequence classes such as Classes 0/I, I, II and III.
These criteria have since been refined by several authors 
(e.g., \citet{Harvey06,Harvey07,Gutermuth08,Gutermuth09,Evans09} and the references therein), 
in order to account for possible contaminations from 
broad-line AGNs, PAH-emitting galaxies, 
shocked emission blobs/knots and unresolved PAH-emission-contaminated apertures, 
which may lead to wrongful identifications of YSOs. We used the updated  
criteria given clearly in the Appendix A of \citet{Gutermuth09} to delineate the PMS sources. 
The total number of point sources
identified that are common to all the four IRAC bands is 3376; of these, 235 sources are found to be 
contaminations (1 PAH galaxy, 6 shocked emissions, 228 PAH aperture-contaminations), while there are 327 
YSOs. 

After removing the contaminants, we used the criteria based on the spectral 
index, $\alpha_{\lambda}$ (= dlog($\lambda$F$_{\lambda}$)/dlog($\lambda$)), 
to classify the YSOs (numbering 327 as shown above) 
into different evolutionary classes (see e.g., \citet{Green94,Smith04,Lada06}).
We followed \citet{Billot10} in the classification of Class 0/I as sources whose $\alpha_{IRAC}$ is 
$>$ -0.3; Class II as those having -0.3 $>$ $\alpha_{IRAC}$ $>$ -1.6; and 
Class III as those having -1.6 $>$ $\alpha_{IRAC}$ $>$ -2.6 (termed as sources with 
faint or anemic disks by \citet{Lada06}). In applying these classifications to M8, 
we have not considered the flat-spectrum sources (e.g., \citet{Green94}) as a separate class
but included them in the Class 0/I, the sources with in-falling envelopes (see \citet{Billot10}). 
The sources with $\alpha_{IRAC}$ $<$ -2.6 are taken as stars with purely photospheric emissions.
With the $\alpha_{IRAC}$ classification, we obtain 64 Class 0/I, 168 Class II sources
and 95 Class III sources. The rest of the sources, numbering 2814, are purely photospheric sources.

We then verified the selected sample of YSOs for possible interstellar  
extinction/reddening bias (see \citet{Muench07} and the references therein).
\citet{Muench07} showed that only for very large values of A$_{v}$ do sources, with $\alpha_{IRAC}$ 
ranges relevant here, get misclassified as YSOs. Such large values are seen only as 
intrinsic for YSOs as may be inferred from the H-K colour. 
Typically, we can identify the bias by comparing the ratio ($N$) of number of Class II sources to 
that of the Class 0/I, for different values of extinction and see if the ratio changes substantially \citep{Guieu09}. 
In the case of M8, the visual extinction and the reddening (defined as the 
ratio of total to selective extinction, R$_{v}$ = A$_{v}$/E(B-V)) varies from region to region; and   
a value of A$_{v}$ = 3.2 was determined towards the Hourglass region 
for standard reddening (R$_{v}$ = 3.1) \citep{Arias06,Tothill08}. 
For the present purpose, however, we compared the ratio $N$, 
for A$_{v}$ values 0.0, 3.2 and 5.0. 
The ratio $N$ remains at 2.63 with a Poisson error of $\pm$ 0.39 for A$_{v}$ 0.0 and 3.2. 
For A$_{v}$ = 5.0, we get $N$ = 2.73$\pm$0.42. 
The value of $N$ obtained here for M8 is comparable 
with that obtained for the Serpens \citep{Harvey06} and North American Nebula \citep{Guieu09} star 
forming regions, indicating the similarity of ages of these regions. 

Fig 2 shows the mid-infrared colour-colour diagram constructed from the 
IRAC photometry for [3.6]-[4.5] vs [5.8]-[8.0] colour. In this diagram we show  
the photospheric sources as black dots, and the Class 0/I, Class II and Class III sources as 
open circles, open triangles and open squares.  
Table 1 lists the Class 0/I sources while Table 2 gives the Class II sources. Also 
included in Table 1 is the source KS 4 (No. 26; classified as YSO by \citet{Arias06}) for 
which IRAC has detections in Chs. 1 \& 2 only.
Where available, the colours [J-H] and [H-K] along with the H magnitude from JHK surveys
are also given in the Tables with references.  

In order to obtain quantitatively the spatial distribution of YSOs, we 
adopted the nearest-neighbour technique (see \citet{Chavarria08,Guieu09,Evans09}). 
We have followed the suggestion of \citet{Casertano85} to obtain  
surface density distribution, without a bias towards over-estimation. We used a 5 arcsec grid to 
compute the surface number density, defined as $\rho_{n} = (n-1)/A_{n}$, where $A_{n}$ is the surface area 
defined by the radial distance $r_{n}$ to the $n$ (= 5) nearest-neighbours.
Fig 3 shows the YSO spatial density contours with the inner and outer contours 
representing 10 and 5 YSOs/pc$^{2}$ respectively. 
The maximum densities are about 20 YSOs/pc$^{2}$. 
Also shown in the figure are the spatial 
distributions of all the 327 YSOs (a) and the contaminants (b). 
One can notice in Fig 3 several small clusters (about 7) 
isolated by the technique used as above.
 These clusters are mostly confined to the Hourglass, NGC 6530, M8E and 
the ridge regions along with two more clumps east of H36.
In order to identify the cluster members as against the isolated or scattered cases, 
we computed the empirical cumulative distribution as 
a function of nearest-neighbour distance and found 
that the sources within an arbitrarily chosen inflection 
distance $d_{c}$ (that signifies the maximum separation between the cluster members) 
of $\sim$ 0.75 pc (0.03$^{o}$ at 1.25 kpc). By varying $d_{c}$ 
between 0.65 and 0.85 pc, the number of cluster members does not change significantly. 
The Class 0/I and Class II YSOs in clusters constitute about 60\% of the total number detected
and are confined mainly within the YSO density contours (corresponding to about 7 clusters) 
shown in Fig 3. In comparison, only about 26 \% of the Class III sources 
(having ``anemic'' disks) occur in the clusters.   
The ratio $N$ mentioned before varies from cluster to cluster with an average of 2.29 which 
is comparable to the one for the entire sample (2.63) within the Poisson errors. 
It may be noted here that the surface density of YSOs derived by us is very 
close to the values obtained for star forming regions 
elsewhere (see \citet{Billot10} and references therein). 

It may be noted here that since M8 ($l$ = 5.958; $b$ = -1.167) is located near 
the mid-plane of the Galaxy, there exists a possibilty of our YSO sample being still contaminated 
from other intrinsically ``red sources", such as AGB stars. 
Recently \citet{Robitaille08} prepared an extensive catalogue of such red sources based on the {\it Spitzer} 
GLIMPSE and MIPSGAL surveys. While the best way to distinguish between 
YSOs and AGB stars is by spectroscopy, these authors
showed that the two classes are well separated in the [8.0-24.0] colour space, 
YSOs being redder than AGB stars in this space (see also \citet{Whitney08}). 
Since in the present case of M8 we do not have 
the 24 $\mu$m data (from MIPS), we have estimated the AGB contamination 
by using the criteria based on the IRAC magnitudes and colour space \citep{Robitaille08}. 
Since AGB stars are unlikely to occur in clusters, we have removed the clustered YSOs from our estimates. 
We find that our Class 0/I and II samples may be contaminated by AGB stars up to about 19\%,
while we do not find any contamination for our Class III sources. As pointed out by 
\citet{Robitaille08}, these separation criteria (including [8.0-24.0] colour) 
are ``only approximate and there is likely to be contamination in both directions".

Fig 4 shows Ch3 (5.8 $\mu$m) image overlaid by the IRAC Class 0/I (open circles) 
and Class II (open triangles) sources identified by using the $\alpha_{IRAC}$ criteria (see Tables 1 and 2).
Also shown (as black star symbols) in the figure are 
the positions of sub-mm (850$\mu$m) gas clumps (taken from Table 1 of \citet{Tothill08}). 
The YSO density contours (in white) are also shown in the figure. 
A number of the Class 0/I and Class II sources occur 
very close to the dense filamentary/pillar-like 
structures seen all along the ridge region in the east-west direction 
as well as perpendicular to it in the central ridge or NGC 6530 region (see Fig 1). 
A majority of these sources are found to be present in the vicinity of the 
sub-mm gas clumps (marked by black star symbols in Fig 4).   
As mentioned earlier, M8E is a young compact high-mass star forming region. 
The central source of this cluster was resolved into a protostar M8E-IR and
M8E-Radio, a B2 type star that is responsible for the compact HII region \citep{Simon84}. 
While IRAC bands are saturated for M8E-IR itself, about 6 Class I or flat-spectrum  
and 10 Class II sources are identified by IRAC in a region of 4 arcmin$^{2}$ around M8E.
These sources are not common with those listed in Table 6 of \citet{Tothill08}.
We did not find any IRAC pre-main-sequence sources in the vicinity of the source IRAS 18014-2428, 
believed to be another young star forming region (corresponding to the sub-mm gas clump
called SE3 \citep{Tothill02}).

Triggered star formation by the ``collect and collapse'' process \citep{Elmegreen77}
could be responsible for the existence of the IRAC sources, possibly started by the stellar winds or 
expanding HII regions associated with the nearby massive stars, 
viz. HD 165052, M8E, HD 164806, HD 164816, 9 Sgr and Her 36 
(see \citet{Tothill02}). While the ionization fronts from 9 Sgr and Her 36 
have expanded well into the molecular cloud, that from M8E seems to have started later, as is evident 
from the bright narrow rim in front of M8E seen in Fig 1. Thus the small cluster in M8E region 
may be younger than others in the region. \citet{Linz09} modelled the spectral energy distribution 
of M8E-IR and concluded that it is a B0 type YSO. That the star formation is sequential in M8 starting 
from north-west regions to southern edge regions has been shown by \citet{Damiani04}, based on Chandra X-ray 
survey of NGC 6530 region and its neighbourhood in conjunction with 2MASS data 
and optical surveys (see \citet{Tothill08} for a discussion).  
 
\subsection{IRAC Ratio Maps}
As mentioned earlier, the IRAC bands contain a number of prominent molecular lines/features. 
Ch1 contains H$_2$ vibrational-rotational lines while Ch2-4 mostly contain pure rotational lines. 
Ch1, 3 and 4 also contain the PAH features at 3.3, 6.2, 7.7 and 8.6 $\mu$m; 
but Ch2 does not include any PAH features. In addition to these molecular lines/features,
IRAC bands also contain hydrogen recombination lines, notably 
the Br$\alpha$ line (4.05 $\mu$m) in Ch2, which can be used to trace HII regions. 
Several authors have utilised the ratios of IRAC bands to identify some of the atomic and molecular 
diagnostics mentioned above (e.g., \citet{Smith05,Povich07,Neufeld08}). 
Since it is difficult to assess the contribution of different atomic or molecular transitions to 
different channels, the ratio maps are only indicative; until/unless supplemented by 
spectroscopic evidence. 

The Ch2 is more sensitive to H$_{2}$ lines of high excitation temperatures 
while the Ch4 represents rotational lines of low excitation temperatures \citep{Neufeld08}.
Likewise, the Ch2 does not have any PAH features while Ch4 has.  
Thus, in the ratio image of Ch2/Ch4, the brighter regions indicate emission regions 
from higher excitations from H$_{2}$ and the darker regions indicate PAH emission. 
This trend is reversed in the image of Ch4/Ch2 (i.e., bright regions show PAH emission and 
dark regions the H$_{2}$ emission). However in HII regions, the H recombination 
lines (Br$\alpha$ and Pf$\beta$) are more significant contributors to Ch2 rather than the H$_{2}$ lines. 

In order to make the ratio maps, point sources from 
all the IRAC images are removed by using an extended aperture of 12.2 arcsec 
and sky annulus of 12.2-24.4 arcsec in IRAF/DAOPHOT software \citep{Reach05}. 
Then these residual frames are subjected to median filtering with a width of 15 pixels 
and smoothing by 3$\times$3 pixels using ``boxcar'' algorithm \citep{Povich07}. 
Fig 5a gives the ratio map of Ch2/Ch4  
in a region of 0.67 $\times$ 0.46 pc$^{2}$ around Her 36. Contours overlaid
on the ratio map represent the H$\alpha$ emission observed by HST 
(in F656N filter image extracted from HST public archive). 
The minimum and maximum values of the contours are 5288 and 14100 counts respectively;  
six contours are drawn with interval of 1762 counts.  
In the vicinity of Her 36 the ratio map Ch2/Ch4 shows bright regions coinciding very nicely 
with the Hourglass HII region (as traced out by the contours of H$\alpha$). 
The bright regions coinciding with the Hourglass can not be 
attributed to the molecular hydrogen lines of higher 
excitation temperature present in Ch2 in comparison with those of lower excitation 
temperature present in Ch4 (see \citet{Smith05}). Further, the molecular hydrogen 
(1-0 S(1) at 2.12 $\mu$m) images presented by \citet{Burton02} do not show substantial 
emission around the Hourglass region (refer Figs 3 and 4 of \citet{Burton02}). 
Considering the fact that the IRAC Ch2 also contains 
Hydrogen Br$\alpha$ line (as well as Pf$\beta$ (4.65 $\mu$m)), we may 
attribute the bright regions in the ratio map of Ch2/Ch4 coinciding with Hourglass, 
as due to Br$\alpha$ (and Pf$\beta$) emission. 
One can notice narrow bright regions surrounding Her 36 (and 
the near-IR source KS1) in the ratio map of Ch2/Ch4. 
These regions are coincident with those of molecular hydrogen in \citet{Burton02}. 
In these regions it is likely that the molecular hydrogen transitions 
of high excitation temperature may be responsible rather than Br$\alpha$. 
A similar trend is seen in the ratio image of Ch2/Ch3 also. 
Fig 5b gives the ratio image Ch4/Ch2 in the same region as in Fig 5a. In this figure, 
the ratio contours are overlaid on ratio image for better clarity and insight.
One can notice a bright (corresponding to dark region in Fig 5a) ``cavity''-like 
structure to the east of the Hourglass. Actually, 
this structure occurs towards the east of the regions of very high column 
density reported by \citet{Arias06} towards north and east of the Hourglass. 
It is possible that the PAH molecules can be shielded from high energy UV photons 
by these dense regions. But the molecules may be excited by 
the low energy (non-ionising) UV photons that can escape from 
the narrow dense regions between the Hourglass and the ``cavity''. 
Thus we may attribute the bright tubular structure to PAH emissions.

It may be mentioned here that the ratio image of Ch2/Ch1 does not show the Hourglass as prominently 
as in Ch2/Ch4 or Ch2/Ch3. This could be because of the fact that Ch1 
contains Pf$\gamma$ (3.7 $\mu$m) and Pf$\delta$ (3.3 $\mu$m) lines
which may be cancelling partly the recombination lines in Ch2. 
In fact we do clearly see the Hourglass HII region in the ratio images of Ch1/Ch3 and Ch1/Ch4 also. 
Thus the Ch1 and Ch2 bring out well the H recombination emission in the Hourglass.
Ch4 is unable to depict the HII region in spite of the presence of Pf$\alpha$ (7.5 $\mu$m), 
mainly because of the fact that its spectral response is inferior to that of Ch1 (or that of Ch2) \citep{Smith05}. 
For a HII region under Case B situation \citep{Hummer87} 
with a kinetic temperature of 10$^{4}$ K and electron densities of 10$^{2}$-10$^{6}$ cm$^{-3}$, 
the combined relative intensities of Pf$\gamma$ and Pf$\delta$ are a factor of $\sim$ 1.4 
times lower than that of Pf$\alpha$. Hence the combined detected flux of Pf$\gamma$ 
and Pf$\delta$ in Ch1 can exceed that of Pf$\alpha$ in Ch4. 

Elsewhere in M8 the ratio maps seem to have a different story to tell. 
Fig 6 gives the ratio maps of the ridge regions in M8 complex in an area of 
24.2 x 20.0 arcmin$^{2}$, to the east/south-east of 
Her 36: Ch1/Ch2 (top), Ch3/Ch2 (middle) and Ch4/Ch2 (bottom).
The ratio images show bright rims corresponding to the filamentary 
structures all along and perpendicular to the east-west ridge region. 
These regions seem to be bright in all bands, except Ch2 that is free of PAH features. 
Hence it is tempting to attribute the bright regions seen in the ratio images in Fig 6 to the PAH features. 
It is unlikely that these bright regions are a result of molecular hydrogen emission; since they 
occur in all the bands though at different excitation temperatures (mostly rotational lines) and should have 
cancelled each other in the ratio maps.     

\section{Conclusions}

The important conclusions of this work are as follows:

\begin{enumerate}
\item{{\it Spitzer} IRAC photometry of M8 region revealed 64 Class 0/I and 168 Class II YSOs. About 60\%
of these are present in about 7 small clusters with spatial surface densities of 10-20 YSOs/pc$^{2}$;}
\item{These sources are positioned close to the sub-mm gas clumps and the 
filamentary or pillar like structures present in M8. It is possible that the formation of these 
sources could have been triggered by stellar winds or expanding HII regions associated with 
the massive stars in the region;}
\item{The ratio map Ch2/Ch4 reveals Br$\alpha$ emission corresponding to the Hourglass HII region powered 
by Her 36 and its inverted ratio (Ch4/Ch2) identifies PAH emission in a cavity, east of the Hourglass;}
\item{The ratio maps Ch1/Ch2, Ch3/Ch2 and Ch4/Ch2 indicate the presence of PAH emission in both 
the ridges oriented along E-W and NE-SW directions.}
\end{enumerate}  

\section*{Acknowledgments}

The research work is supported by the Department of Space, Government of India at PRL. 
This work is based (for a large part) on observations made with the 
Spitzer Space Telescope, which is operated by the Jet Propulsion Laboratory, 
California Institute of Technology under a contract with NASA.
We acknowledge the use of data from the 2MASS, 
which is a joint project of the University of Massachusetts and the 
Infrared Processing and Analysis Center/California Institute of Technology, 
funded by the NASA and the NSF. The authors sincerely appreciate the very useful 
comments from the anonymous referee.

\begin{figure*}
\includegraphics[width=\textwidth]{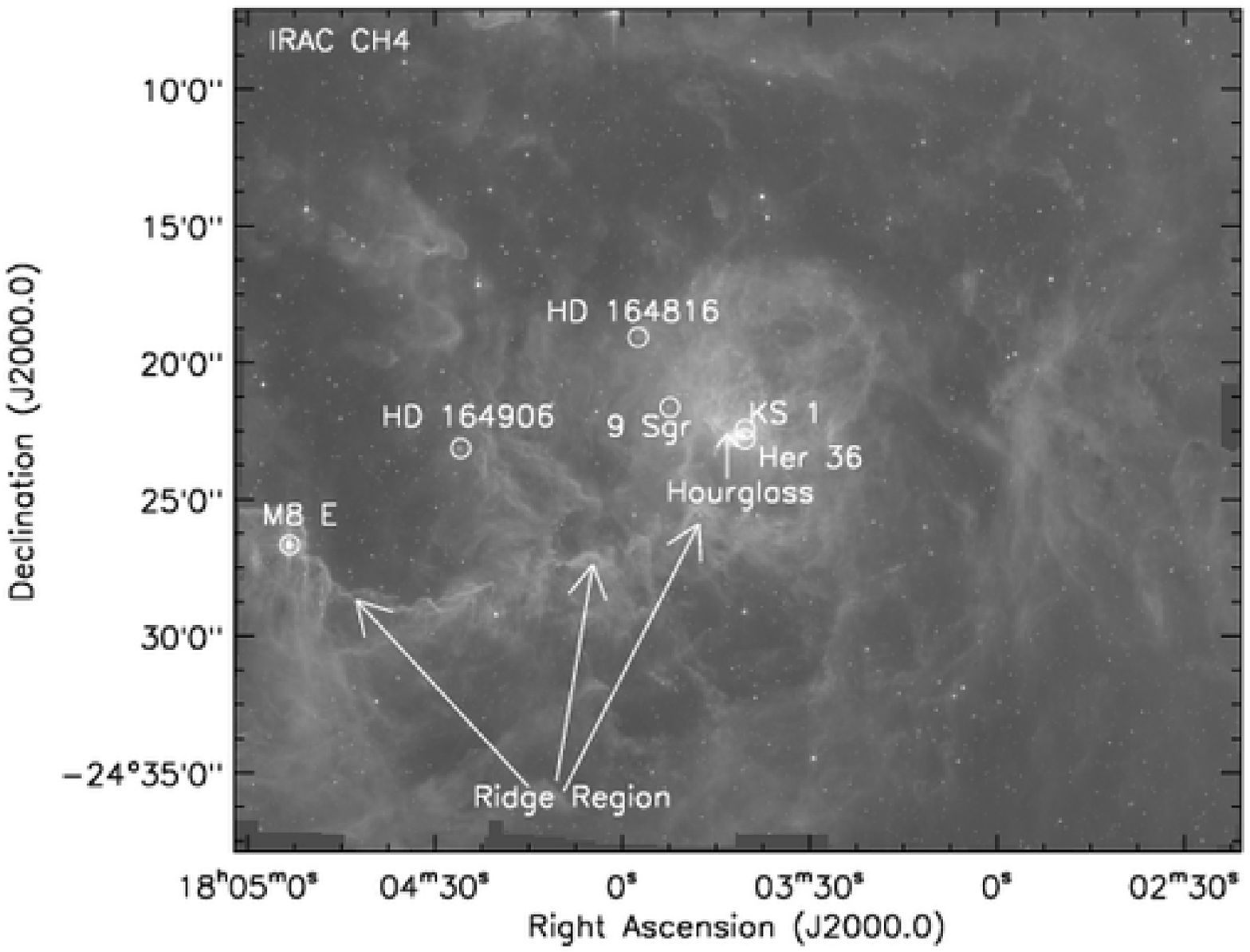}
\caption{Spitzer IRAC Ch4 (8.0 $\mu$m) image of Lagoon Nebula region 
($\sim$ 42.5$\times$30.0 arcmin$^{2}$). The locations
of well known sources are shown circled: the massive O type stars 
Her 36 and 9 Sgr; the early B type stars HD 164906 and HD 164816;  
the near-infrared source KS1 just north of Her 36; and 
the young compact massive star-forming region M8E to the extreme east. 
The Hourglass HII region is shown by the arrow near the core of M8. 
The young cluster NGC 6530 is situated just to the west of the massive star HD 164906.
The arrows at the bottom show the ridge region in the east-west direction. One can 
also notice filamentary features in the central ridge 
running north-south near the star HD 164906. The massive star HD 165052 is towards
the east of M8E but it is not covered by the IRAC observations.}
\label{fig1}
\end{figure*}

\begin{figure*}
\centering
\includegraphics[width=\textwidth]{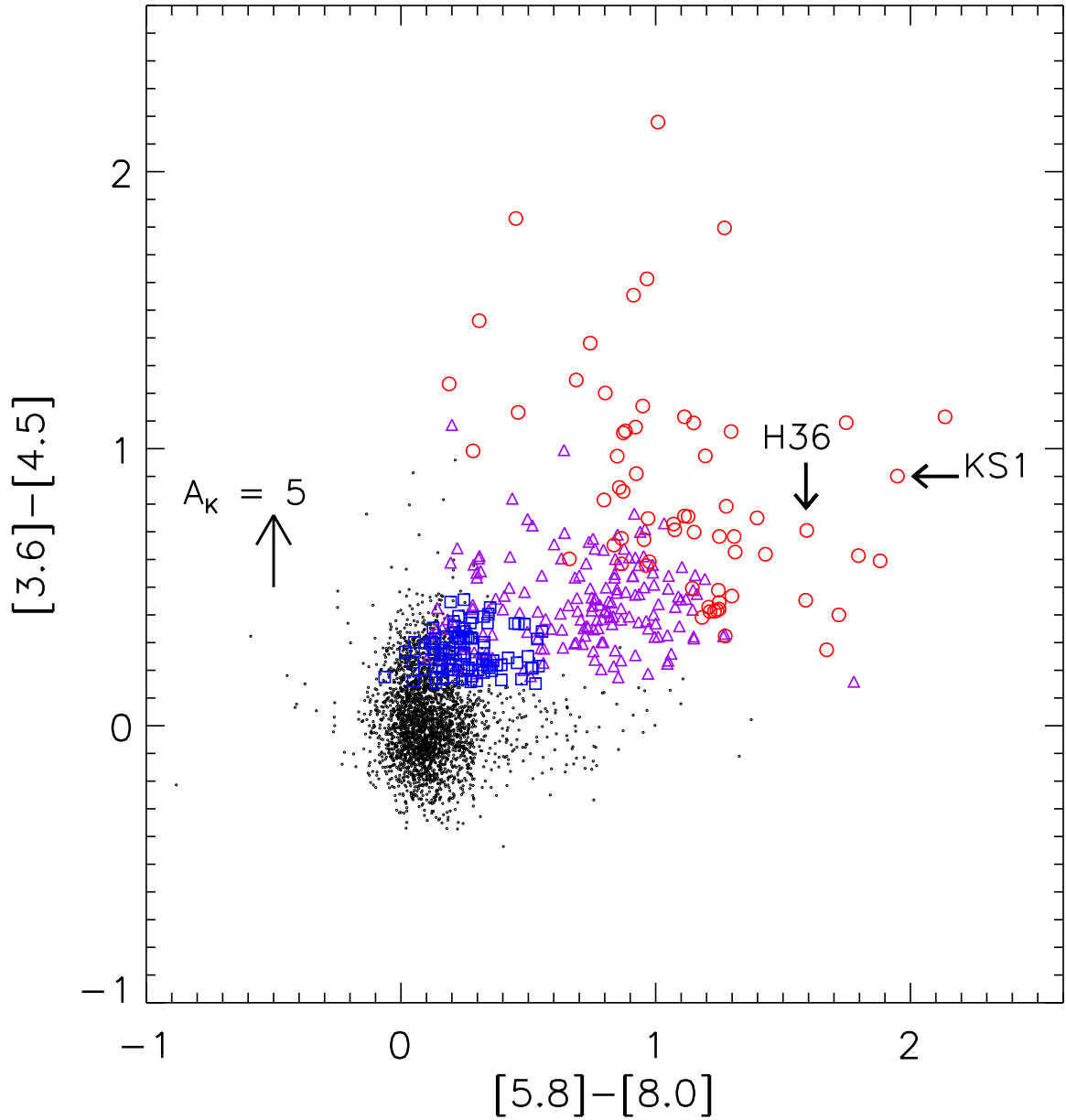}
\caption{Mid-IR color-color diagram using the Spitzer IRAC bands for all the sources identified 
within the region shown in Fig 1. The sources KS1 and Her 36 are marked by arrows. The upward arrow 
on the left shows the extinction vector for A$_{K}$ = 5 mag, using average extinction law from 
\citet{Flaherty07}. The black dots around the centre (0,0) locate 
the stars with only photospheric emissions. 
The open squares (blue), open triangles (violet) and open circles (red) 
represent respectively, Class III, Class II and Class 0/I sources, obtained from the $\alpha_{IRAC}$ criteria.}
\label{fig2}
\end{figure*}

\begin{figure*}
\centering
\subfigure[]{
\includegraphics[width=0.55\textwidth]{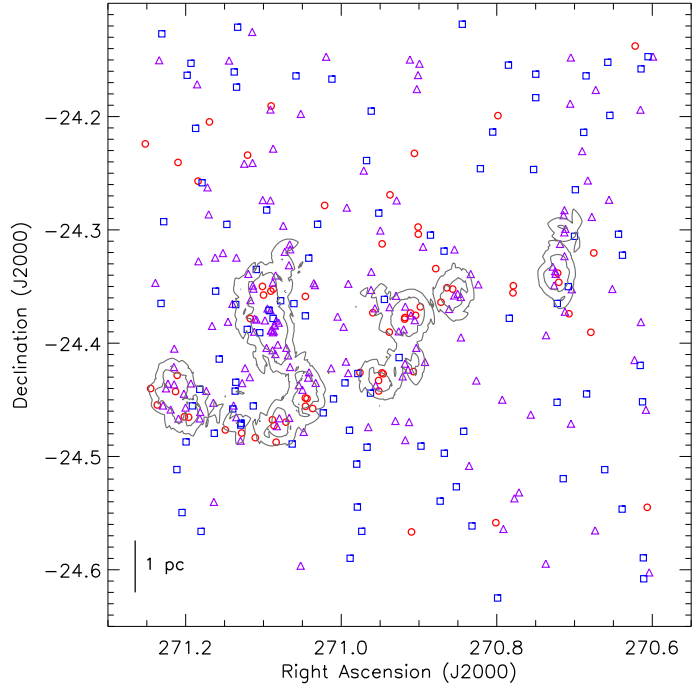}}
\subfigure[]{
\includegraphics[width=0.55\textwidth]{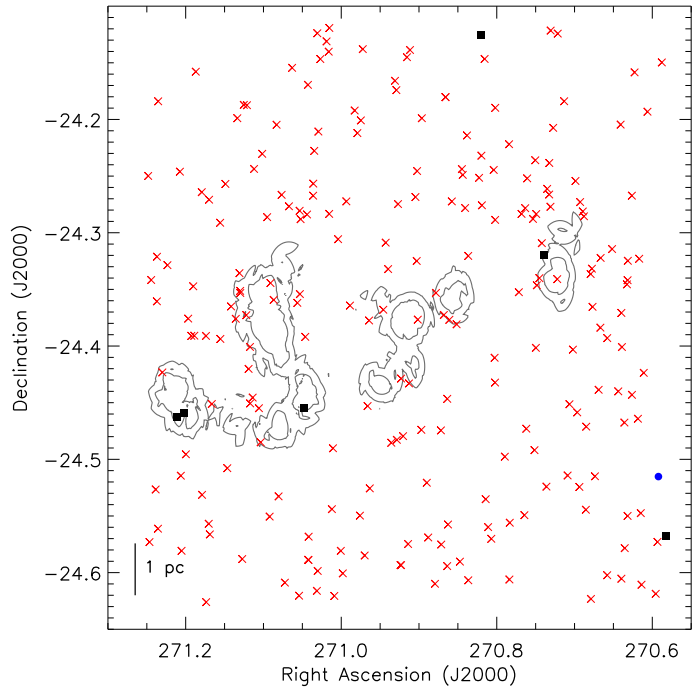}}
\caption{a: Spitzer IRAC field of M8 showing all the 327 YSOs: open circles (red) show Class 0/I sources, 
open triangles (violet) show Class II and open squares (blue) show Class III sources;  
b: Spitzer IRAC field of M8 showing contaminants: crosses (red) show unresolved PAH aperture 
contaminations, filled squares (black) show shocked emission knots and the 
lone filled circle (blue) the PAH galaxy contaminant.
In both the figures, the contours show YSO iso-density at 5 (outer) and 10 (inner) YSOs/pc$^{2}$.} 
\label{fig3}
\end{figure*}

\begin{figure*}
\centering
\includegraphics[width=\textwidth]{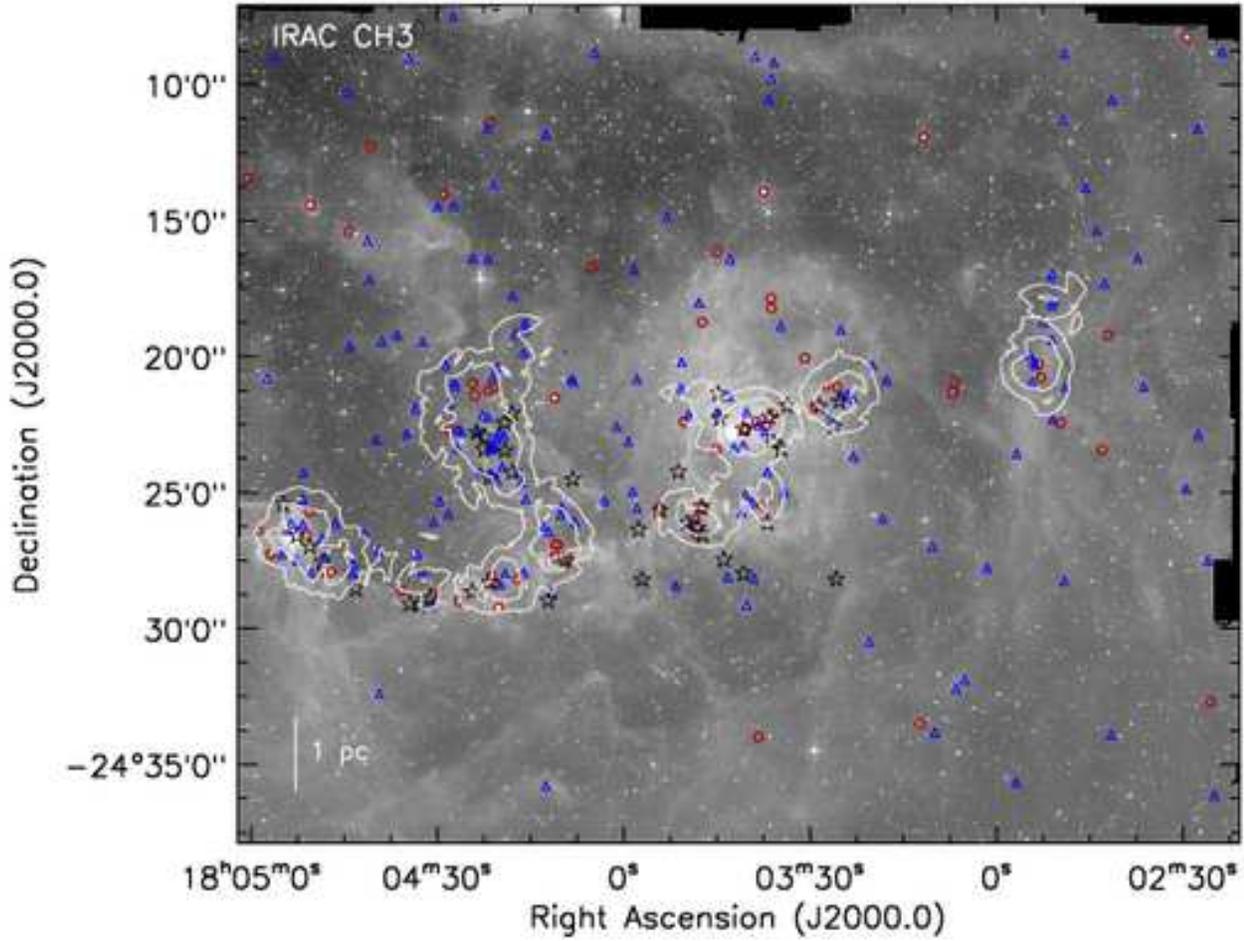}
\caption{Spitzer IRAC Ch3 (5.8 $\mu$m) image of M8 
($\sim$ 42.5$\times$30.0 arcmin$^{2}$) superposed by IRAC Class 0/I and II sources 
and sub-mm gas clumps. The open circles (red) and open triangles (blue) show the IRAC 
Class 0/I and II sources respectively (from Tables 1 and 2 of the present work); 
the black star symbols represent the locations of the sub-mm gas clumps (from Table 1 of \citet{Tothill08}).
The overlaid contours (white) are YSO density contours generated using a grid size of 5 arcsec: 
the inner contours are 10 YSOs/pc$^{2}$ and the outer contours represent 5 YSOs/pc$^{2}$.}
\label{fig4}
\end{figure*}

\begin{figure*}
\centering
\subfigure[]{
\includegraphics[width=0.75\textwidth]{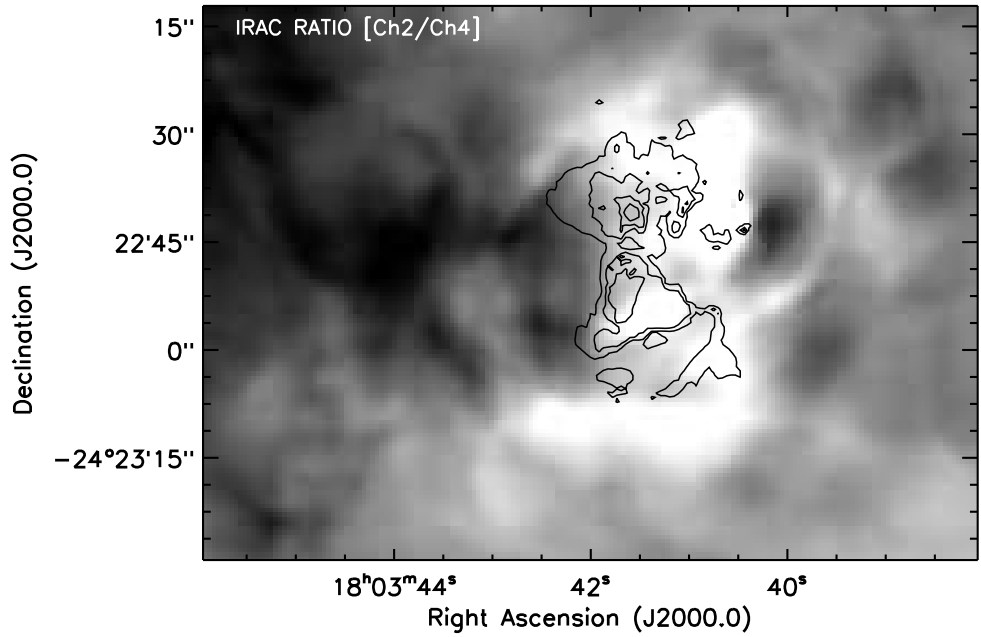}}
\subfigure[]{
\includegraphics[width=0.75\textwidth]{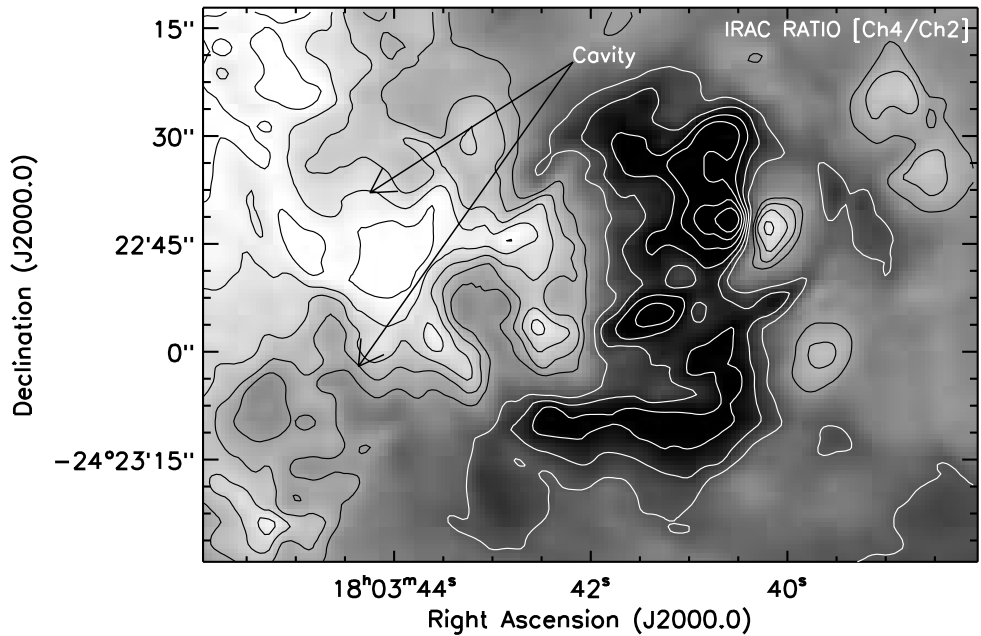}}
\caption{a: Ratio image of Ch2/Ch4 of M8 (Lagoon Nebula) in log scale overlaid by HST H$\alpha$ (F656N filter) 
contours, in a region around Her 36 of size 110$\times$76 arcsec$^{2}$. 
The contour levels are between 5288 and 14100 counts. 
The bright regions indicate the prominence of Ch2 over Ch4, while the 
dark regions have the reverse trend. The bright ``3'' shaped structure is more extended than
the H$\alpha$ contours and nearly coincide with them.
b: Ratio image of Ch4/Ch2 of M8 (Lagoon Nebula) in log scale. The bright regions indicate the prominence of Ch4 over Ch2.
For better insight, the ratio image is overlaid by the ratio contours: the black contours
in the bright regions represent ratio levels between 17.4 and 22.8; and the white contours
in the dark regions indicate the ratio levels of 14.3 to 8.4.
The region of possible PAH emission is the bright tubular structure within the ``cavity'',
seen east of the Hourglass.} 
\label{fig5}
\end{figure*}

\begin{figure*}
\centering
\includegraphics[width=0.55\textwidth]{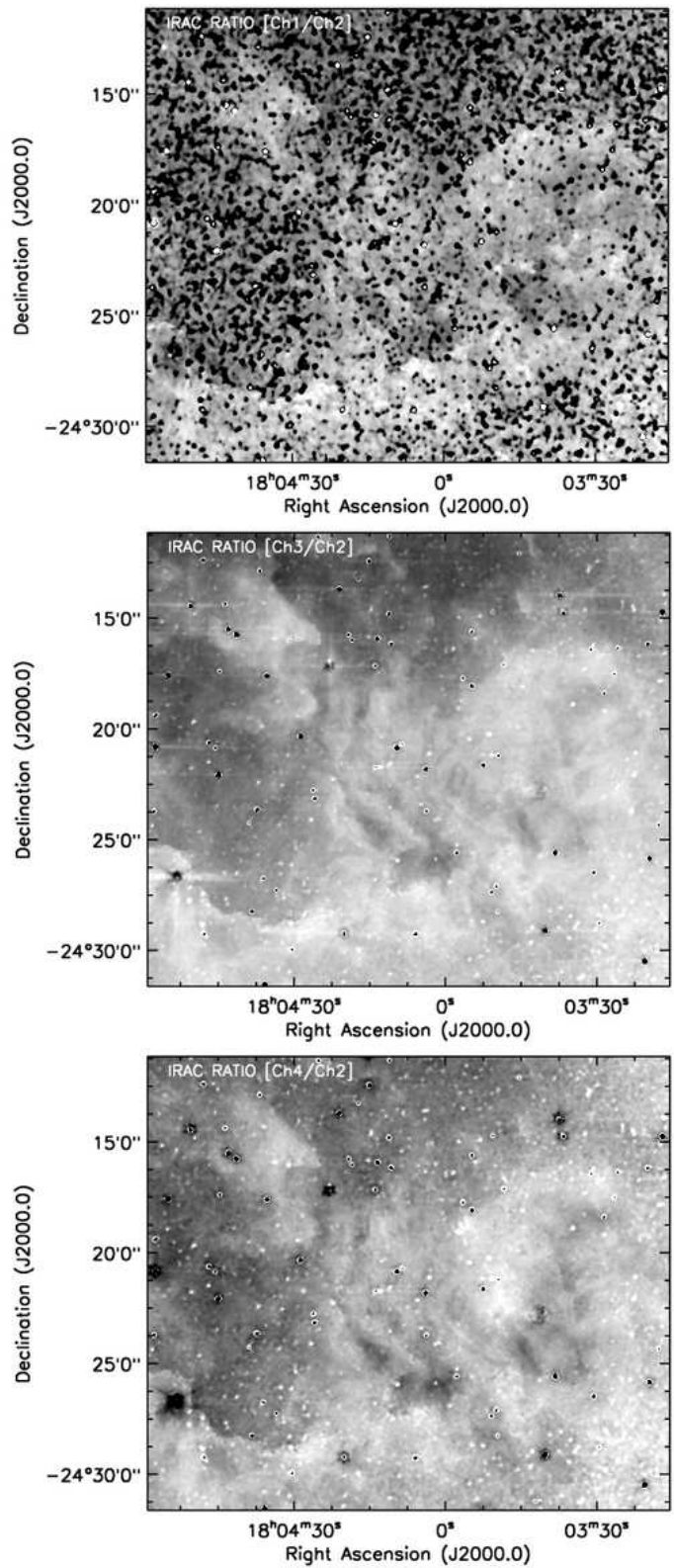}
\caption{Ratio maps of Ch1/Ch2 (top), Ch3/Ch2 (middle) and Ch4/Ch2 (bottom) in 
log scale (in a region of size $\sim$ 24.2 $\times$ 20.0 arcmin$^{2}$),
showing the ridges and filamentary structures (towards east/south-east of Her 36). The dominance of Ch1, Ch3 and 
Ch4 (bright portions) over Ch2 in the ridge region and filamentary regions can be noticed, which 
may be attributed to the PAH emission in these regions. The black dots (seen mostly in Ch1/Ch2) 
and the white dots (seen mostly in Ch3/Ch2 and Ch4/Ch2) are the result of the residueing process.} 
\label{fig6}
\end{figure*}

\end{document}